\documentclass[11pt,a4paper]{article}
\usepackage{amsmath, amsfonts, amssymb}
\usepackage{a4wide}
\usepackage{parskip}
\usepackage{enumitem}
\usepackage{xcolor}

\usepackage{hyperref}
\usepackage{booktabs}
\usepackage{caption}
\usepackage{siunitx}
\usepackage{tabularx}
\usepackage{graphicx}
\usepackage{adjustbox}
\usepackage{multirow}
\usepackage{amsthm}
\usepackage{bm}
\usepackage{lscape}
\usepackage{float}
\usepackage[noadjust]{cite}

\def\qed{\hfill {$\square$}\goodbreak \medskip}

\newtheorem{theorem}{Theorem}[section]
\newtheorem{lemma}[theorem]{Lemma}
\theoremstyle{definition}
\newtheorem{definition}[theorem]{Definition}
\newtheorem{example}[theorem]{Example}

\theoremstyle{remark}
\newtheorem{remark}[theorem]{Remark}

\numberwithin{equation}{section}

\newcommand{\Tr}{\textnormal{Tr}}
\newcommand{\Supp}{\textnormal{Supp}}

\usepackage{tikz,xcolor,hyperref}

\definecolor{lime}{HTML}{A6CE39}
\DeclareRobustCommand{\orcidicon}{%
	\begin{tikzpicture}
		\draw[lime, fill=lime] (0,0) 
		circle [radius=0.16] 
		node[white] {{\fontfamily{qag}\selectfont \tiny ID}};
		\draw[white, fill=white] (-0.0625,0.095) 
		circle [radius=0.007];
	\end{tikzpicture}
	\hspace{-2mm}
}

\foreach \x in {A, ..., Z}{%
	\expandafter\xdef\csname orcid\x\endcsname{\noexpand\href{https://orcid.org/\csname orcidauthor\x\endcsname}{\noexpand\orcidicon}}
}

\begin{document}
	\date{}
	{\vspace{0.01in}
		\title{ Subfield codes of $C_D$-codes over $\mathbb{F}_2[x]/\langle x^3-x \rangle$ are really nice!}
		\author{{\bf Anuj Kumar Bhagat\footnote{email: {\tt anujkumarbhagat632@gmail.com}}\orcidA{},\;\bf Ritumoni Sarma\footnote{	email: {\tt ritumoni407@gmail.com}}\orcidB{}\; and \bf Vidya Sagar\footnote{email: {\tt vsagariitd@gmail.com}}\orcidC{}} \\ Department of Mathematics,\\ Indian Institute of Technology Delhi,\\Hauz Khas, New Delhi-110016, India. }
		\maketitle
		\begin{abstract} A non-zero $\mathbb{F}$-linear map from a finite-dimensional commutative $\mathbb{F}$-algebra to $\mathbb{F}$ is called an $\mathbb{F}$-valued trace if its kernel does not contain any non-zero ideals. In this article, we utilize an $\mathbb{F}_2$-valued trace of the $\mathbb{F}_2$-algebra $\mathcal{R}_2:=\mathbb{F}_2[x]/\langle x^3-x\rangle$ to study binary subfield code $\mathcal{C}_D^{(2)}$ of $\mathcal{C}_D:=\{\left(x\cdot d\right)_{d\in D}: x\in \mathcal{R}_2^m\}$ for each defining set $D$ derived from a certain simplicial complex. For $m\in \mathbb{N}$ and $X\subseteq \{1, 2, \dots, m\}$, define $\Delta_X:=\{v\in \mathbb{F}_2^m: \Supp(v)\subseteq X\}$ and $D:=(1+u^2)D_1+u^2D_2+(u+u^2)D_3,$ a subset of $\mathcal{R}_2^m,$ where $u=x+\langle x^3-x\rangle, D_1\in \{\Delta_L, \Delta_L^c\},\, D_2\in \{\Delta_M, \Delta_M^c\}$ and $ D_3\in \{\Delta_N, \Delta_N^c\}$, for $L, M, N\subseteq \{1, 2, \dots, m\}.$ The parameters and the Hamming weight distribution of the binary subfield code $\mathcal{C}_D^{(2)}$ of $\mathcal{C}_D$ are determined for each $D.$ These binary subfield codes are minimal under certain  mild conditions on the cardinalities of $L, M$ and $N$. Moreover, most of these codes  are distance-optimal. Consequently, we obtain a few infinite families of minimal, self-orthogonal and distance-optimal binary linear codes that are either $2$-weight or $4$-weight. It is worth mentioning that we have obtained several new distance-optimal binary linear codes.

			\medskip
			
			
			\medskip
			
			
		\end{abstract}
\section{Introduction}\label{Section 1}
Given a code $\mathcal{C}$ over $\mathbb{F}_{q^m},$ the computation of the subfield code $\mathcal{C}^{(q)}$ of $\mathcal{C}$ gets simplified due to the existence of the trace map from $\mathbb{F}_{q^m}$ to $\mathbb{F}_q$ (for instance, see \cite{ding2019subfield}). In \cite{sagar2022certain}, codes over $\mathbb{F}_2$ are obtained using a non-zero functional (which is not a trace map) from codes defined over an $\mathbb{F}_2$-algebra; these are called ``subfield-like codes''. However, a suitable choice (which may not exist) of an $\mathbb{F}_q$-functional of an $\mathbb{F}_q$-algebra, the subfield-like code of a given code is the subfield code. In fact, every such suitable choice of an $\mathbb{F}_q$-functional of an $\mathbb{F}_q$-algebra is a trace map of the $\mathbb{F}_q$-algebra. For instance, see \cite{BHAGAT2024102360} for the existence and construction of an $\mathbb{F}_q$-valued trace of a finite-dimensional $\mathbb{F}_q$-algebra. 
\par
 Simplicial complexes got massive attention in algebraic coding theory due to the work of Chang et al. in \cite{chang2018linear}, where they first studied linear codes using simplicial complexes. Since then, many linear codes and their corresponding subfield codes have been studied using simplicial complexes (see  \cite{heng2020two, hyun2020infinite, liu2023linear, sagar2022certain, sagarE, sagarI, sagar2023octanary, shi2022few, wu2022quaternary, wu2019optimal}). Linear codes whose defining set is obtained from a simplicial complex were shown to have nice parameters; for instance, in \cite{hyun2020infinite}, Hyun et al. produced infinite families of optimal binary linear codes constructed using simplicial complexes.
\par
To calculate the error-correcting capability and the error probability of error-detection of a linear code, its weight distribution is required (see \cite{ding2015class}, \cite{dinh2015recent}, \cite{peterson1961cyclic}) and as such, determining the weight distribution of a linear code has become a hot research topic. Moreover, to maximize the error-correcting capability of a linear code, its Hamming distance should be maximized. As a result, finding distance-optimal linear codes is one of the central research topics in algebraic coding theory. The weight distribution of linear codes over $\mathbb{F}_{2^k}$ and their corresponding subfield codes whose defining set is derived from a simplicial complex is obtained in \cite{liu2023linear}.
\par
A non-zero codeword in a linear code is called minimal if its support does not contain the support of any codeword different from its nonzero scalar multiples. The covering problem of a linear code is to determine the set of all minimal codewords of the code. In general, the covering problem of linear codes is hard \cite{huffman2021concise}. Minimal linear codes are special because all their non-zero codewords are minimal. Minimal linear codes are important because of their applications in secret sharing schemes (see \cite{shamir1979share}, \cite{yuan2005secret}) and in secure two-party computation \cite{chabanne2013towards}. In secret sharing schemes based on minimal linear codes, the access structure of the secret sharing scheme is completely determined\cite{ding2003covering}. In secure two-party computation, minimal linear codes ensure privacy in the protocol for the secure two-party computation. Furthermore, minimal linear codes are also important as they can be decoded using the minimum distance decoding rule\cite{ashikhmin1998minimal}. Because of the easy decoding and many applications of minimal linear codes, an active research topic is to produce minimal linear codes.
\par
In \cite{sagar2022certain}, the authors study linear codes over $\mathcal{R}_2:=\mathbb{F}_2[x]/\langle x^3-x \rangle$ whose defining set is obtained from certain simplicial complexes. They produced certain minimal binary codes using a Gray image and subfield-like codes using an $\mathbb{F}_2$-linear functional. Hence, a natural endeavour is to explore the construction of the subfield codes $\mathcal{C}_D^{(2)}$ of $\mathcal{C}_D$, where the defining set $D$ is obtained using a simplicial complex.
\par
 In this article, we study the subfield code of $\mathcal{C}_D:=\{\left(x\cdot d\right)_{d\in D}: x\in \mathcal{R}_2^m\},$ where $m\in \mathbb{N}$, $\mathcal{R}_2:=\mathbb{F}_2[x]/\langle x^3-x\rangle$ and $D\subseteq \mathcal{R}_2^m$, a defining set (of $\mathcal{C}_D$) constructed using a simplicial complex. In fact, we obtain the Hamming weight distribution of the subfield code $\mathcal{C}_D^{(2)}$ and sufficient conditions for it to be a minimal code. We also show that some of these codes meet the Griesmer bound, and some other are distance-optimal although they do not meet the Griesmer bound. Furthermore, we produce a few infinite families of minimal, self-orthogonal and distance-optimal binary linear codes that are either $2$-weight or $4$-weight. To the best of our knowledge, this is the first article on the construction of subfield codes of codes over an $\mathbb{F}_2$-algebra; in fact, this is the first instance to compute (binary) subfield codes of codes over an $\mathbb{F}_2$-algebra using an $\mathbb{F}_2$-valued trace.
\par
The remaining sections of this article are as follows. Definitions and preliminaries are presented in the next section. In Section \ref{Section 3}, we construct subfield codes of certain linear codes over $\mathcal{R}_2,$ and in Section \ref{Section 4}, we find the parameters and the weight distribution of the subfield code $\mathcal{C}^{(2)}_D$. In Section \ref{Section 5}, we list some results of recent work in the direction of our work. We also present a table containing several new optimal codes as a subclass of our obtained codes in this section. Section \ref{Section 6} concludes the article.
\section{Definitions and Preliminaries}\label{Section 2}
Throughout this article, $\mathbb{F}_q$ denotes the finite field with $q$ elements and we denote the set $\{1,2,\dots, m\}$ by $[m].$ For $v=(v_1,\dots, v_n)\in \mathbb{F}_q^n,$ the subset
$\Supp(v):=\{i\in [m]: v_i\neq 0 \}$ of $[m]$ is called the \textit{support of v} and $wt:=|Supp(v)|$ is called the \textit{Hamming weight of $v.$} An \textit{$[n,k,d]$-linear code} $\mathcal{C}$ of length $n$ is a $k$-dimensional $\mathbb{F}_q$-subspace of the vector space $\mathbb{F}_q^n,$ where $d:=\min\{wt(c): 0\neq c\in \mathcal{C}\}.$ Elements of $\mathcal{C}$ are called \textit{codewords}. The string $(1,A_1,\dots,A_n)$ is called the \textit{(Hamming) weight distribution} of $\mathcal{C},$ where $A_i:=|\{c\in \mathcal{C}: wt(c)=i\}|.$ Moreover, if the total number of $i\ge 1$ such that $A_i\neq 0$ is $l,$ then $\mathcal{C}$ is called an \textit{$l$-weight linear code}.
\par
For $u,v\in \mathbb{F}_q^n,$ we say that $u$ \textit{covers} $v$ (denoted by $v\preceq u$) if $\Supp(u)\supseteq\Supp(v).$ A non-zero codeword $u\in \mathcal{C}$ is called \textit{minimal} if $v\preceq u, v\in \mathcal{C}\setminus\{0\},$ then $v=\lambda u$ for some $\lambda\in \mathbb{F}_q^*$ and if every non-zero codeword of $\mathcal{C}$ is minimal, then $\mathcal{C}$ is called a \textit{minimal code}. In \cite{ashikhmin1998minimal}, Ashikhmin-Barg gave a sufficient condition for linear code over $\mathbb{F}_q$ to be minimal, which is as follows:
\begin{lemma}\cite{ashikhmin1998minimal}
    Let $\mathcal{C}$ be a linear code over $\mathbb{F}_q$ with minimum and maximum non-zero (Hamming) weights $wt_{min}$ and $wt_{max}$ respectively. Then
    \begin{equation*}
        \frac{wt_{min}}{wt_{max}}>\frac{q-1}{q} \implies \mathcal{C}\, \text{is minimal}.
    \end{equation*}
\end{lemma}
\begin{lemma}[Griesmer Bound] \cite{griesmer1960bound}
    If $\mathcal{C}$ is an $[n,k,d]$-linear code over $\mathbb{F}_q,$ then
    \begin{equation}\label{Griesmer bound}
        \sum_{i=0}^{k-1} \left\lceil \frac{d}{q^i}\right\rceil\le n,
    \end{equation}
    where $\lceil\cdot\rceil$ denotes the ceiling function.
\end{lemma}
\begin{definition}
    A linear code is called a \textit{Griesmer code} if the equality holds in Equation (\ref{Griesmer bound}).
\end{definition}
\begin{definition}
    An $[n,k,d]$-linear code over $\mathbb{F}_q$ is called \textit{distance-optimal} if there is no $[n,k,d+1]$-linear code over $\mathbb{F}_q.$
\end{definition}
\begin{remark}
    Every Griesmer code is distance-optimal.
\end{remark}
We next recall a sufficient condition for binary linear codes to be Euclidean self-orthogonal.
\begin{theorem}\cite{huffman2010fundamentals}
    If the Hamming weight of every non-zero codeword of a binary linear code is divisible by $4,$ then the code is Euclidean self-orthogonal.
\end{theorem}
Let $D=\{\{d_1< d_2<...<d_n\}\}$ be an ordered multiset, where each $d_i\in \mathcal{R}_2^m.$ Define
$$
\mathcal{C}_D:=\{\left(x\cdot d_1,\, x\cdot d_2,\, \dots,\, x\cdot d_n\right): x\in \mathcal{R}_2^m\},
$$
where $x\cdot y:=\sum_{j=0}^{m} x_iy_i$ for $x=(x_1,\dots, x_m), y=(y_1,\dots, y_m)\in \mathcal{R}_2^m.$ Then $\mathcal{C}_D$ is a linear code of length $n$ over $\mathcal{R}_2$ and we call the ordered multiset $D$ to be the \emph{defining set} of the code $\mathcal{C}_D.$
\begin{definition}\cite{generalized}
    Let $S$ be a ring (need not be commutative) with unity and $R$ be a subring of $S$ sharing the same unity. Then a homomorphism of left $R$-modules $\Tr_R^S: S\to R$ is called an \textit{$R$-valued trace} of $S$ if it is surjective and $\ker(\Tr_R^S)$ does not contain any non-zero left ideals of $S.$
\end{definition}
\subsection{Simplicial complexes}
The map $f:\mathbb{F}_2^{m}\to 2^{[m]}$ given by $f(v)=\Supp(v)$ is a bijection. Henceforth, we write $v$ instead of $\Supp(v)$ whenever needed. A subset $\Delta$ of $\mathbb{F}_2^m$ is called a \textit{simplicial complex} if $v\in \Delta$ and $w\preceq v,$ then $w\in \Delta.$ An element $v\in \Delta$ is called \textit{maximal} if for $w\in \Delta,$ $v\preceq w$ imply $v=w.$ For $L\subseteq[m],$ the set $\Delta_L:=\{v\in \mathbb{F}_2^m: \Supp(v) \subseteq L\}$ is called the \textit{simplicial complex generated by $L.$} The maximal element of $\Delta_L$ is $f^{-1}(L)$ and $|\Delta_L|=2^{|L|}.$
\par
For a subset $Q$ of $\mathbb{F}_2^{m},$ define the $m$-variable generating function \cite{chang2018linear} $\mathcal{H}_Q(y_1,y_2,\dots,y_m)$ by
\begin{equation}
    \mathcal{H}_Q(y_1,y_2,\dots,y_m):=\sum_{v\in Q}\prod_{i=1}^{m}y_i^{v_i}\in \mathbb{Z}[y_1,y_2,\dots, y_m],
\end{equation}
where $v=(v_1, v_2, \dots v_m)$ and $\mathbb{Z}$ denotes the ring of integers.
\begin{lemma}\cite{chang2018linear}
    Suppose $\Delta\subseteq \mathbb{F}_2^{m}$ is a simplicial complex. If $\mathcal{F}$ is the collection of all maximal elements of $\Delta,$ then 
    \begin{equation}
        \mathcal{H}_Q(y_1,y_2,\dots,y_m)=\sum_{\phi\neq S\subseteq \mathcal{F}}(-1)^{|S|+1}\prod_{i\in \cap S}(1+y_i)
    \end{equation}
    where $\cap S=\bigcap_{F\in S} \Supp(F).$ In particular, we have
    \begin{equation}
        |\Delta|=\sum_{\phi\neq S\subseteq \mathcal{F}}(-1)^{|S|+1}2^{|\cap S|}.
    \end{equation}
\end{lemma}
\section{Construction of subfield codes of linear codes over $\mathcal{R}_2$}\label{Section 3}
Let $\mathcal{R}_q$ be a finite-dimensional commutative $\mathbb{F}_q$-algebra that possesses an $\mathbb{F}_q$-valued trace $\tau.$ Suppose that $\mathcal{C}$ is a linear code of length $n$ over $\mathcal{R}_q$ generated by a matrix $G$ and $\mathcal{B}$ is a basis for $\mathcal{R}_q$ over $\mathbb{F}_q.$ The code $\mathcal{C}^{(q)}$ over $\mathbb{F}_q$ generated by the matrix which is obtained by replacing each entry of $G$ by its column representation relative to $\mathcal{B},$ is called a \emph{subfield code} of $\mathcal{C}$. Now we have the following Theorem from \cite{BHAGAT2024102360}.

\begin{theorem}\label{subfield theorem}\cite{BHAGAT2024102360}
    Suppose $\mathcal{B}=\{\bm{\alpha}_1,\dots,\bm{\alpha}_m\}$ be a basis for $\mathcal{R}_q$ over $\mathbb{F}_q$ and let $\mathcal{C}$ be linear code of length $n$ over $\mathcal{R}_q$ generated by:
    $$
    G=\begin{bmatrix}
        \bm{g}_{11} & \bm{g}_{12} & \cdots & \bm{g}_{1n} \\
        \bm{g}_{21} & \bm{g}_{22} & \cdots & \bm{g}_{2n} \\
        \vdots  & \vdots & \ddots & \vdots\\
        \bm{g}_{k1} & \bm{g}_{k2} & \cdots & \bm{g}_{kn} \\
    \end{bmatrix}$$
    Then the subfield code $\mathcal{C}^{(q)}$ of $\mathcal{C}$ is generated by
    $$
        G^{(q)}=\begin{bmatrix}
            G_1^{(q)}\\
            G_2^{(q)}\\
            \vdots\\
            G_k^{(q)}
        \end{bmatrix},
    $$
    where $\tau:\mathcal{R}_q\to \mathbb{F}_q$ is an $\mathbb{F}_q$-valued trace of $\mathcal{R}_q$ and for $1\leq i\leq k,$
    $$
        G_i^{(q)}=\begin{bmatrix}
            \tau(\bm{g}_{i1}\bm{\alpha}_1) & \tau(\bm{g}_{i2}\bm{\alpha}_1) & \dots & \tau(\bm{g}_{in}\bm{\alpha}_1)\\
            \tau(\bm{g}_{i1}\bm{\alpha}_2) & \tau(\bm{g}_{i2}\bm{\alpha}_2) & \dots & \tau(\bm{g}_{in}\bm{\alpha}_2)\\
            \vdots & \vdots & \ddots & \vdots\\
            \tau(\bm{g}_{i1}\bm{\alpha}_m) & \tau(\bm{g}_{i2}\bm{\alpha}_m) & \dots & \tau(\bm{g}_{in}\bm{\alpha}_m)\\
        \end{bmatrix}.
    $$
\end{theorem}
Set $q=2$ and consider the $\mathbb{F}_2$-algebra $\mathcal{R}_2:=\mathbb{F}_2[x]/\langle x^3-x\rangle.$ Then the map $\tau:\mathcal{R}_2\to\mathbb{F}_2$ given by $a+bu+cu^2\mapsto c$ is an $\mathbb{F}_2$-valued trace of $\mathcal{R}_2$ by \cite{BHAGAT2024102360}, where $u=x+\langle x^3-x \rangle$. Fix the ordered basis $\mathcal{B}=\{\bm{e}_1=1+u^2<\bm{e}_2=u^2<\bm{e}_3=u+u^2\}$ for $\mathcal{R}_2$ over $\mathbb{F}_2,$. Let $\bm{g}_{ij}=g_{ij}^{(1)}\bm{e}_1+g_{ij}^{(2)}\bm{e}_2+g_{ij}^{(3)}\bm{e}_3 \in \mathcal{R}_2,$ where $g_{ij}^{(1)}, g_{ij}^{(2)}, g_{ij}^{(3)} \in \mathbb{F}_2.$ Then
\begin{align*}
    \tau(\bm{g}_{ij}\bm{e}_1)&=g_{ij}^{(1)},\\
    \tau(\bm{g}_{ij}\bm{e}_2)&=g_{ij}^{(2)}+g_{ij}^{(3)},\\
    \tau(\bm{g}_{ij}\bm{e}_3)&=g_{ij}^{(2)}.
\end{align*}
Now, by Theorem \ref{subfield theorem}, we have the following result.
\begin{theorem}\label{subfield code over R_2}
    Let $\mathcal{B}=\{\bm{e}_1=1+u^2<\bm{e}_2=u^2<\bm{e}_3=u+u^2\}$ be an ordered basis for $\mathcal{R}_2$ over $\mathbb{F}_2.$ Suppose $G:=\bm{e}_1G_1+\bm{e}_2G_2+\bm{e}_3G_3\in M_{k\times n}(\mathcal{R}_2)$ generates the linear code $\mathcal{C}$ over $\mathcal{R}_2,$ where $G_i \in M_{k\times n}(\mathbb{F}_2), 1\leq i\leq 3.$ Then the subfield code $\mathcal{C}^{(2)}$ of $\mathcal{C}$ is linear over $\mathbb{F}_2$ and is generated by 
    \begin{equation}
        G^{(2)}=\begin{bmatrix}
                    G_1\\
                    G_2+G_3\\
                    G_2
                \end{bmatrix}.
    \end{equation}
    Moreover, $\mathcal{C}_D^{(2)}=\mathcal{C}_{D^{(2)}}$, that is, if the defining set of the code $\mathcal{C}_D$ is $D=\bm{e}_1D_1+\bm{e}_2D_2+\bm{e}_3D_3\subseteq\mathcal{R}_2^m,$ where $D_i\subseteq \mathbb{F}_2^m,$ $1\leq i\leq 3,$ then the defining set of $\mathcal{C}_D^{(2)}$ is
    \begin{equation}
        D^{(2)}=\{(d_1, d_2+d_3, d_2): d_i\in D_i, 1\leq i\leq 3\}.
    \end{equation}
\end{theorem}
\begin{proof}
    Immediate from the above discussion. \qed
\end{proof}
Observe that the map $c_{D}^{(2)}:({\mathbb{F}_2^m})^3\to \mathcal{C}_{D}^{(2)}\subseteq (\mathbb{F}_2)^{\vert D^{(2)}\vert }$ defined by 
\begin{equation}
    c_{D}^{(2)}(\bm{v})=(\bm{v}\cdot\bm{d})_{\bm{d}\in D^{(2)}}
\end{equation}
is a surjective linear transformation of $\mathbb{F}_2$-vector spaces. By Theorem \ref{subfield code over R_2},
 \begin{equation}
     \mathcal{C}_D^{(2)}:=\{c_D^{(2)}(\alpha,\beta, \gamma)=(\alpha, \beta, \gamma)\cdot(d_1, d_2+d_3, d_2)_{d_i\in D_i}: \alpha,\beta,\gamma \in \mathbb{F}_2^m\}.
 \end{equation}
 \par
 Now, we have
 \begin{equation}\label{weight}
     \begin{split}
         wt(c_{D}^{(2)})&=wt\left(((\alpha, \beta,\gamma)\cdot(d_1, d_2+d_3, d_2))_{d_1\in D_1,\, d_2\in D_2,\, d_3\in D_3}\right)\\    
     &=wt\left((\alpha d_1+(\beta+\gamma)d_2+\beta d_3)_{d_1\in D_1,\, d_2\in D_2,\, d_3\in D_3}\right)\\
     &=|D|-\frac{1}{2}\sum_{d_1\in D_1}\sum_{d_2\in D_2}\sum_{d_3\in D_3}(1+(-1)^{\alpha d_1+(\beta+\gamma)d_2+\beta d_3})\\
     &=\frac{|D|}{2}-\frac{1}{2}\sum_{d_1\in D_1}(-1)^{\alpha d_1}\sum_{d_2\in D_2}(-1)^{(\beta+\gamma)d_2}\sum_{d_3\in D_3}(-1)^{\beta d_3}
     \end{split}
 \end{equation}
Now, for $\alpha\in \mathbb{F}_2^m$ and $\phi\neq L\subseteq [m],$ if we define a boolean function $\varphi(\alpha|L)$ as follows
\begin{equation}
    \varphi(\alpha|L):=\prod_{i\in L}(1-\alpha_i)=\begin{cases} 
      0, & \text{if}\,\, \alpha\cap L\neq \phi \\
      1, & \text{if}\,\, \alpha\cap L= \phi 
   \end{cases}
\end{equation}
then we have
\begin{equation}\label{boolean}
    \begin{split}
        \sum_{t\in \Delta_L}(-1)^{\alpha t}&=\mathcal{H}_{\Delta_L}((-1)^{\alpha_1},(-1)^{\alpha_2},\dots,(-1)^{\alpha_m})\\
    &=\prod_{i\in L}(1+(-1)^{\alpha_i})=\prod_{i\in L}(2-2\alpha_i)\\
    &=2^{|L|}\prod_{i\in L}(1-\alpha_i)=2^{|L|}\varphi(\alpha|L).
    \end{split}
\end{equation}
 \begin{lemma}\cite{shi2022few}\label{complement formula}
     For $\alpha\in \mathbb{F}_2^m,$ we have
     \begin{equation*}
         \sum_{t\in \Delta_L^c}(-1)^{\alpha t}=2^m\delta_{0,\alpha}-2^{|L|}\varphi(\alpha|L),
     \end{equation*}
     where $\delta$ denotes the Kronecker delta function and $\Delta_{L}^c=\mathbb{F}_2^m\setminus\Delta_L.$
 \end{lemma}
\section{Weight distributions of binary subfield code $\mathcal{C}_D^{(2)}$ of $\mathcal{C}_D$ using simplicial complexes}\label{Section 4}
In this section, we study the binary subfield code $\mathcal{C}_D^{(2)}$ of $\mathcal{C}_D$ over $\mathcal{R}_2$ for various choices of $D.$
 \begin{theorem}\label{main theorem}
    Suppose $[m]\in \mathbb{N}$ and $L, M, N \subseteq [m].$
    \begin{enumerate}
        \item\label{main theorem 1} Let $D=\bm{e}_1\Delta_L+\bm{e}_2\Delta_M+\bm{e}_3\Delta_N\subseteq \mathcal{R}_2^m.$ Then $\mathcal{C}_D^{(2)}$ is a $1$-weight binary linear code with parameters $[2^{|L|+|M|+|N|},|L|+|M|+|N|, 2^{|L|+|M|+|N|-1}].$ The Hamming weight distribution of $\mathcal{C}_D^{(2)}$ is presented in Table \ref{tab:1}. Moreover, $\mathcal{C}_D^{(2)}$ meets the Griesmer bound and hence it is distance-optimal.
        \item\label{main theorem 2} Let $D=\bm{e}_1\Delta_L^c+\bm{e}_2\Delta_M+\bm{e}_3\Delta_N\subseteq \mathcal{R}_2^m.$ Then $\mathcal{C}_D^{(2)}$ is a $2$-weight binary linear code with parameters $[(2^m-2^{|L|})\times 2^{|M|+|N|},m+|M|+|N|, (2^m-2^{|L|})\times 2^{|M|+|N|-1}].$ The Hamming weight distribution of $\mathcal{C}_D^{(2)}$ is presented in Table \ref{tab:2}. Moreover, $\mathcal{C}_D^{(2)}$ meets the Griesmer bound and hence it is distance-optimal.
        \item\label{main theorem 3} Let $D=\bm{e}_1\Delta_L+\bm{e}_2\Delta_M^c+\bm{e}_3\Delta_N\subseteq \mathcal{R}_2^m.$ Then $\mathcal{C}_D^{(2)}$ is a $2$-weight binary linear code with parameters $[(2^m-2^{|M|})\times 2^{|L|+|N|},m+|L|+|N|, (2^m-2^{|M|})\times 2^{|L|+|N|-1}].$ The Hamming weight distribution of $\mathcal{C}_D^{(2)}$ is presented in Table \ref{tab:3}. Moreover, $\mathcal{C}_D^{(2)}$ meets the Griesmer bound and hence it is distance-optimal.
        \item\label{main theorem 4} Let $D=\bm{e}_1\Delta_L+\bm{e}_2\Delta_M+\bm{e}_3\Delta_N^c\subseteq \mathcal{R}_2^m.$ Then $\mathcal{C}_D^{(2)}$ is a $2$-weight binary linear code with parameters $[(2^m-2^{|N|})\times 2^{|L|+|M|},m+|L|+|M|, (2^m-2^{|N|})\times 2^{|L|+|M|-1}].$ The Hamming weight distribution of $\mathcal{C}_D^{(2)}$ is presented in Table \ref{tab:4}. Moreover, $\mathcal{C}_D^{(2)}$ meets the Griesmer bound and hence it is distance-optimal.
        \item\label{main theorem 5} Let $D=\bm{e}_1\Delta_L^c+\bm{e}_2\Delta_M^c+\bm{e}_3\Delta_N\subseteq \mathcal{R}_2^m.$ Then $\mathcal{C}_D^{(2)}$ is a $4$-weight binary linear code with parameters $[(2^m-2^{|L|})\times (2^m-2^{|M|})\times 2^{|N|}, 2m+|N|, (2^m-2^{|L|}-2^{|M|})\times 2^{m+|N|-1}].$ The Hamming weight distribution of $\mathcal{C}_D^{(2)}$ is presented in Table \ref{tab:5}. Moreover, $\mathcal{C}_D^{(2)}$ is distance-optimal if $2^{|L|+|M|+|N|}\le 2(m-1)+|N|.$
        \item\label{main theorem 6} Let $D=\bm{e}_1\Delta_L^c+\bm{e}_2\Delta_M+\bm{e}_3\Delta_N^c\subseteq \mathcal{R}_2^m.$ Then $\mathcal{C}_D^{(2)}$ is a $4$-weight binary linear code with parameters $[(2^m-2^{|L|})\times (2^m-2^{|N|})\times 2^{|M|}, 2m+|M|, (2^m-2^{|L|}-2^{|N|})\times 2^{m+|M|-1}].$ The Hamming weight distribution of $\mathcal{C}_D^{(2)}$ is presented in Table \ref{tab:6}. Moreover, $\mathcal{C}_D^{(2)}$ is distance-optimal if $2^{|L|+|M|+|N|}\le 2(m-1)+|M|.$
        \item\label{main theorem 7} Let $D=\bm{e}_1\Delta_L+\bm{e}_2\Delta_M^c+\bm{e}_3\Delta_N^c\subseteq \mathcal{R}_2^m.$ Then $\mathcal{C}_D^{(2)}$ is a $4$-weight binary linear code with parameters $[(2^m-2^{|M|})\times (2^m-2^{|N|})\times 2^{|L|}, 2m+|L|, (2^m-2^{|M|}-2^{|N|})\times 2^{m+|L|-1}].$ The Hamming weight distribution of $\mathcal{C}_D^{(2)}$ is presented in Table \ref{tab:7}. Moreover, $\mathcal{C}_D^{(2)}$ is distance-optimal if $2^{|L|+|M|+|N|}\le 2(m-1)+|L|.$
        \item\label{main theorem 8} Let $D=\bm{e}_1\Delta_L^c+\bm{e}_2\Delta_M^c+\bm{e}_3\Delta_N^c\subseteq \mathcal{R}_2^m.$ Then $\mathcal{C}_D^{(2)}$ is a $8$-weight binary linear code with parameters $[(2^m-2^{|L|})\times (2^m-2^{|M|})\times (2^{M}-2^{|N|}), 3m, 2^{-1}\times(2^m-2^{|L|})\times(2^m-2^{|M|})\times (2^{M}-2^{|N|})-(2^m-2^{\min\{|L|, |M|, |N|\}})\times2^{|L|+|M|+|N|-\min\{|L|, |M|, |N|\}-1}].$ The Hamming weight distribution of $\mathcal{C}_D^{(2)}$ is presented in Table \ref{tab:8}.
        \item\label{main theorem 9} Let $D=\bm{e}_1\Delta_L+\bm{e}_2\Delta_M+\bm{e}_3\Delta_N\subseteq \mathcal{R}_2^m.$ Then $\mathcal{C}_{D^{(c)}}^{(2)}$ is a $2$-weight binary linear code with parameters $[(2^{3m}-2^{|L|+|M|+|N|}), 3m, 2^{3m-1}-2^{|L|+|M|+|N|-1}].$ The Hamming weight distribution of $\mathcal{C}_{D^{(c)}}^{(2)}$ is presented in Table \ref{tab:9}. Moreover, $\mathcal{C}_{D^{(c)}}^{(2)}$ meets the Griesmer bound and hence it is distance-optimal.
    \end{enumerate}
\end{theorem}
\begin{proof}
    We prove part $(8)$. Observe that $\mathcal{C}_D^{(2)}$ has length $|D^{(2)}|=|D|=(2^m-2^{|L|})\times (2^m-2^{|M|})\times (2^{M}-2^{|N|}).$ Now, from Equations \eqref{weight}, Equation \eqref{boolean} and Lemma \ref{complement formula}, we have
    \begin{multline*}
        wt(c_D^{(2)}(\alpha,\beta, \gamma))=\frac{|D|}{2}-\frac{1}{2}\times(2^m\delta_{0,\alpha}-2^{|L|}\varphi(\alpha|L))\times(2^m\delta_{0,\beta+\gamma}-2^{|M|}\varphi(\beta+\gamma|M))\\\times(2^m\delta_{0,\beta}-2^{|N|}\varphi(\beta|N)).
    \end{multline*}
    Case (1): $wt(c_D^{(2)}(\alpha,\beta, \gamma))=0 \iff \alpha=0, \beta+\gamma=0, \beta=0.$\\
    In this case, $\#\alpha=1, \#\beta=1, \#\gamma=1.$ Therefore, $\#(\alpha,\beta,\gamma)=1.$

    \noindent
    Case (2): $\alpha\neq 0, \beta+\gamma=0, \beta=0.$ This implies $\gamma=0.$ In this case,
    \begin{equation*}
        wt(c_D^{(2)}(\alpha,\beta, \gamma))=\frac{|D|}{2}+2^{|L|-1}\times(2^m-2^{|M|})\times(2^m-2^{|N|})\varphi(\alpha|L).
    \end{equation*}
    \begin{itemize}
        \item If $\varphi(\alpha|L)=0,$ then $wt(c_D^{(2)}(\alpha,\beta, \gamma))=\frac{|D|}{2}.$
        In this case,\\
            $\#\alpha=(2^{|L|}-1)\times 2^{m-|L|},\, \#\beta=1,\, \#\gamma=1.$\\
        Therefore, $\#(\alpha,\beta,\gamma)=(2^{|L|}-1)\times 2^{m-|L|}=2^m-2^{m-|L|}.$
        \item If $\varphi(\alpha|L)=1,$ then $wt(c_D^{(2)}(\alpha,\beta, \gamma))=\frac{|D|}{2}+2^{|L|-1}\times(2^m-2^{|M|})\times(2^m-2^{|N|}).$ In this case,\\
            $\#\alpha=2^{m-|L|}-1,\,\#\beta=1,\,\#\gamma=1$.\\
        Therefore, $\#(\alpha,\beta,\gamma)=2^{m-|L|}-1.$
    \end{itemize}
    
\noindent
    Case (3): $\alpha=0, \beta+\gamma=0, \beta\neq 0.$ This implies $\gamma=\beta\neq 0.$ In this case,
    \begin{equation*}
        wt(c_D^{(2)}(\alpha,\beta, \gamma))=\frac{|D|}{2}+2^{|N|-1}\times(2^m-2^{|L|})\times(2^m-2^{|M|})\varphi(\beta|N).
    \end{equation*}
    \begin{itemize}
        \item If $\varphi(\beta|N)=0,$ then $wt(c_D^{(2)}(\alpha,\beta, \gamma))=\frac{|D|}{2}.$
        In this case,\\
            $\#\alpha=1,\,\#\beta=(2^{|N|}-1)\times 2^{m-|N|},\,\#\gamma=1.$\\
        Therefore, $\#(\alpha,\beta,\gamma)=(2^{|N|}-1)\times 2^{m-|N|}=2^m-2^{m-|N|}.$
        \item If $\varphi(\beta|N)=1,$ then $wt(c_D^{(2)}(\alpha,\beta, \gamma))=\frac{|D|}{2}+2^{|N|-1}\times(2^m-2^{|L|})\times(2^m-2^{|M|}).$ In this case,\\
            $\#\alpha=1,\,\#\beta=2^{m-|N|}-1,\,\#\gamma=1$.\\
        Therefore, $\#(\alpha,\beta,\gamma)=2^{m-|N|}-1.$
    \end{itemize}

\noindent
    Case (4): $\alpha=0, \beta+\gamma\neq0, \beta= 0.$ This implies $\gamma\neq 0.$ In this case,
    \begin{equation*}
        wt(c_D^{(2)}(\alpha,\beta, \gamma))=\frac{|D|}{2}+2^{|M|-1}\times(2^m-2^{|L|})\times(2^m-2^{|N|})\varphi(\beta+\gamma|M).
    \end{equation*}
    \begin{itemize}
        \item If $\varphi(\beta+\gamma|M)=0,$ then $wt(c_D^{(2)}(\alpha,\beta, \gamma))=\frac{|D|}{2}.$
        In this case, $\beta\cap M\neq \gamma\cap M\implies \gamma\cap M\neq \phi.$ Hence,\\
            $\#\alpha=1,\,\#\beta=1,\,\#\gamma=(2^{|M|}-1)\times 2^{m-|M|}$.\\
        Therefore, $\#(\alpha,\beta,\gamma)=2^m-2^{m-|M|}.$
        \item If $\varphi(\beta+\gamma|M)=1,$ then $wt(c_D^{(2)}(\alpha,\beta, \gamma))=\frac{|D|}{2}+2^{|M|-1}\times(2^m-2^{|L|})\times(2^m-2^{|N|}).$ In this case, $\beta\cap M= \gamma\cap M\implies \gamma\cap M=\phi.$ Hence,\\
            $\#\alpha=1,\,\#\beta=1,\,\#\gamma=2^{m-|M|}-1$.\\
        Therefore, $\#(\alpha,\beta,\gamma)=2^{m-|M|}-1.$
    \end{itemize}

\noindent
    Case (5): $\alpha\neq 0, \beta+\gamma=0, \beta\neq 0.$ This implies $\gamma=\beta\neq 0.$ In this case,
    \begin{equation*}
        wt(c_D^{(2)}(\alpha,\beta, \gamma))=\frac{|D|}{2}-(2^m-2^{|M|})\times2^{|L|+|N|-1}\varphi(\alpha|L)\varphi(\beta|N).
    \end{equation*}
    \begin{itemize}
        \item If $\varphi(\alpha|L)=0$ and $\varphi(\beta|N)=0,$ then $wt(c_D^{(2)}(\alpha,\beta, \gamma))=\frac{|D|}{2}.$
        In this case,\\
            $\#\alpha=(2^{|L|}-1)\times2^{m-|L|},\,\#\beta=(2^{|N|}-1)\times2^{m-|N|},\,\#\gamma=1$.
        Therefore, $\#(\alpha,\beta,\gamma)=(2^{|L|}-1)\times2^{m-|L|}\times(2^{|N|}-1)\times2^{m-|N|}=2^{2m}-2^{2m-|N|}-2^{2m-|L|}+2^{2m-|L|-|N|}.$
        
        \item If $\varphi(\alpha|L)=1$ and $\varphi(\beta|N)=0,$ then $wt(c_D^{(2)}(\alpha,\beta, \gamma))=\frac{|D|}{2}.$
        In this case,\\
            $\#\alpha=2^{m-|L|}-1,\,\#\beta=(2^{|N|}-1)\times2^{m-|N|},\,\#\gamma=1$.\\
        Therefore, $\#(\alpha,\beta,\gamma)=(2^{m-|L|}-1)\times(2^{|N|}-1)\times2^{m-|N|}=2^{2m-|L|}-2^{2m-|L|-|N|}-2^m+2^{m-|N|}.$

        \item If $\varphi(\alpha|L)=0$ and $\varphi(\beta|N)=1,$ then $wt(c_D^{(2)}(\alpha,\beta, \gamma))=\frac{|D|}{2}.$
        In this case,\\
            $\#\alpha=(2^{|L|}-1)\times2^{m-|L|},\,\#\beta=2^{m-|N|}-1,\,\#\gamma=1$.\\
        Therefore, $\#(\alpha,\beta,\gamma)=(2^{|L|}-1)\times2^{m-|L|}\times(2^{m-|N|}-1)=2^{2m-|N|}-2^m-2^{2m-|L|-|N|}+2^{m-|L|}.$

        \item If $\varphi(\alpha|L)=1$ and $\varphi(\beta|N)=1,$ then $wt(c_D^{(2)}(\alpha,\beta, \gamma))=\frac{|D|}{2}-(2^m-2^{|M|})\times2^{|L|+|N|-1}.$
        In this case,
            $\#\alpha=2^{m-|L|}-1,\,\#\beta=2^{m-|N|}-1,\,\#\gamma=1$.\\
        Therefore, $\#(\alpha,\beta,\gamma)=(2^{m-|L|}-1)\times(2^{m-|N|}-1)=2^{2m-|L|-|N|}-2^{m-|L|}-2^{m-|N|}+1.$
    \end{itemize}

\noindent
    Case (6): $\alpha\neq 0, \beta+\gamma\neq 0, \beta=0.$ This implies $\gamma\neq 0.$ In this case,
    \begin{equation*}
        wt(c_D^{(2)}(\alpha,\beta, \gamma))=\frac{|D|}{2}-(2^m-2^{|N|})\times2^{|L|+|M|-1}\varphi(\alpha|L)\varphi(\beta+\gamma|M).
    \end{equation*}
    \begin{itemize}
        \item If $\varphi(\alpha|L)=0$ and $\varphi(\beta+\gamma|M)=0,$ then $wt(c_D^{(2)}(\alpha,\beta, \gamma))=\frac{|D|}{2}.$
        In this case,\\
            $\#\alpha=(2^{|L|}-1)\times2^{m-|L|},\,\#\beta=1,\,\#\gamma=(2^{|M|}-1)\times2^{m-|M|}$.\\
        Therefore, $\#(\alpha,\beta,\gamma)=(2^{|L|}-1)\times2^{m-|L|}\times(2^{|M|}-1)\times2^{m-|M|}=2^{2m}-2^{2m-|M|}-2^{2m-|L|}+2^{2m-|L|-|M|}.$

        \item If $\varphi(\alpha|L)=1$ and $\varphi(\beta+\gamma|M)=0,$ then $wt(c_D^{(2)}(\alpha,\beta, \gamma))=\frac{|D|}{2}.$
        In this case,\\
            $\#\alpha=2^{m-|L|}-1,\,\#\beta=1,\,\#\gamma=(2^{|M|}-1)\times2^{m-|M|}$.\\
        Therefore, $\#(\alpha,\beta,\gamma)=(2^{m-|L|}-1)\times(2^{|M|}-1)\times2^{m-|M|}=2^{2m-|L|}-2^{2m-|L|-|M|}-2^m+2^{m-|M|}.$

        \item If $\varphi(\alpha|L)=0$ and $\varphi(\beta+\gamma|M)=1,$ then $wt(c_D^{(2)}(\alpha,\beta, \gamma))=\frac{|D|}{2}.$
        In this case,\\
            $\#\alpha=(2^{|L|}-1)\times2^{m-|L|},\,\#\beta=1,\,\#\gamma=2^{m-|M|}-1$.\\
        Therefore, $\#(\alpha,\beta,\gamma)=(2^{|L|}-1)\times2^{m-|L|}\times(2^{m-|M|}-1)=2^{2m-|M|}-2^m-2^{2m-|L|-|M|}+2^{m-|L|}.$

        \item If $\varphi(\alpha|L)=1$ and $\varphi(\beta+\gamma|M)=1,$ then $wt(c_D^{(2)}(\alpha,\beta, \gamma))=\frac{|D|}{2}-(2^m-2^{|N|})\times2^{|L|+|M|-1}.$
        In this case,\\
            $\#\alpha=2^{m-|L|}-1,\,\#\beta=1,\,\#\gamma=2^{m-|M|}-1$.\\
        Therefore, $\#(\alpha,\beta,\gamma)=(2^{m-|L|}-1)\times(2^{m-|M|}-1)=2^{2m-|L|-|M|}-2^{m-|L|}-2^{m-|M|}+1.$
    \end{itemize}

\noindent
    Case (7): $\alpha=0, \beta+\gamma\neq 0, \beta\neq0.$ In this case,
    \begin{equation*}
        wt(c_D^{(2)}(\alpha,\beta, \gamma))=\frac{|D|}{2}-(2^m-2^{|L|})\times2^{|M|+|N|-1}\varphi(\beta+\gamma|M)\varphi(\beta|N).
    \end{equation*}
    Subcase (1): $\gamma=0$
    \begin{itemize}
        \item If $\varphi(\beta+\gamma|M)=0$ and $\varphi(\beta|N)=0,$ then $wt(c_D^{(2)}(\alpha,\beta, \gamma))=\frac{|D|}{2}.$
        In this case,\\
            $\#\alpha=1,\,
            \#\beta=(2^{|M|}-1)\times2^{m-|M|}+(2^{|N|}-1)\times 2^{m-|N|}-(2^{|M\cup N|}-1)\times2^{m-|M\cup N|},\,
            \#\gamma=1$.\\
        Therefore, $\#(\alpha,\beta,\gamma)=(2^{|M|}-1)\times2^{m-|M|}+(2^{|N|}-1)\times 2^{m-|N|}-(2^{|M\cup N|}-1)\times2^{m-|M\cup N|}=2^{m}-2^{m-|M|}-2^{m-|N|}+2^{m-|M\cup N|}.$

        \item If $\varphi(\beta+\gamma|M)=1$ and $\varphi(\beta|N)=0,$ then $wt(c_D^{(2)}(\alpha,\beta, \gamma))=\frac{|D|}{2}.$
        In this case,\\
            $\#\alpha=1,\,\#\beta=(2^{|N\setminus M|}-1)\times 2^{m-|M\cup N|},\,\#\gamma=1$.\\
        Therefore, $\#(\alpha,\beta,\gamma)=(2^{|N\setminus M|}-1)\times 2^{m-|M\cup N|}=2^{m-|M|}-2^{m-|M\cup N|}.$

        \item If $\varphi(\beta+\gamma|M)=0$ and $\varphi(\beta|N)=1,$ then $wt(c_D^{(2)}(\alpha,\beta, \gamma))=\frac{|D|}{2}.$
        In this case,\\
            $\#\alpha=1,\,\#\beta=(2^{|M\setminus N|}-1)\times 2^{m-|M\cup N|},\,\#\gamma=1$.\\
        Therefore, $\#(\alpha,\beta,\gamma)=(2^{|M\setminus N|}-1)\times 2^{m-|M\cup N|}=2^{m-|N|}-2^{m-|M\cup N|}.$

        \item If $\varphi(\beta+\gamma|M)=1$ and $\varphi(\beta|N)=1,$ then $wt(c_D^{(2)}(\alpha,\beta, \gamma))=\frac{|D|}{2}-(2^m-2^{|L|})\times 2^{|M|+|N|-1}.$
        In this case,\\
            $\#\alpha=1,\,\#\beta=2^{m-|M\cup N|}-1,\,\#\gamma=1$.\\
        Therefore, $\#(\alpha,\beta,\gamma)=2^{m-|M\cup N|}-1.$
    \end{itemize}

Subcase (2): $\gamma\neq \beta$
    \begin{itemize}
        \item If $\varphi(\beta+\gamma|M)=0$ and $\varphi(\beta|N)=0,$ then $wt(c_D^{(2)}(\alpha,\beta, \gamma))=\frac{|D|}{2}.$\\
        \begin{itemize}
            \item [(i)] $\gamma\cap M=\phi.$
            In this case,\\
                $\#\alpha=1\\
                \#\beta=(2^{|M|}-1)\times2^{m-|M|}+(2^{|N|}-1)\times 2^{m-|N|}-(2^{|M\cup N|}-1)\times2^{m-|M\cup N|}\\
                \#\gamma=2^{m-|M|}-1$.
            \item[(ii)]  $\gamma\cap M\neq\phi.$
            \begin{itemize}
                \item[(a)] $\beta \cap M=\phi$.       
                In this case,\\
                    $\#\alpha=1\\
                    \#\beta=(2^{|N\setminus M|}-1)\times2^{m-|M\cup N|}\\
                    \#\gamma=(2^{|M|}-1)\times 2^{m-|M|}$.
                \item[(b)]$\beta \cap M\neq \phi$. In this case,\\
                    $\#\alpha=1\\
                    \#\beta=(2^{|M|}-1)\times2^{m-|M|}+(2^{|N|}-1)\times 2^{m-|N|}-(2^{|M\cup N|}-1)\times2^{m-|M\cup N|}\\
                    \#\gamma=(2^{|M|}-2)\times 2^{m-|M|}$.
            \end{itemize}
        \end{itemize}
        Therefore, $\#(\alpha,\beta,\gamma)=2^{2m}+2^{2m-|M|-|N|}-2^{2m-|M|}-2^{2m-|N|}-2^{m}+2^{m-|M|}+2^{m-|N|}-2^{m-|M\cup N|}.$

        \item If $\varphi(\beta+\gamma|M)=1$ and $\varphi(\beta|N)=0,$ then $wt(c_D^{(2)}(\alpha,\beta, \gamma))=\frac{|D|}{2}.$\\
        \begin{itemize}
            \item [(i)] $\gamma\cap M=\phi.$
            In this case,\\
            $\#\alpha=1\\
            \#\beta=(2^{|N\setminus M|}-1)\times2^{m-|M\cup N|}\\
            \#\gamma=2^{m-|M|}-2$.
            \item[(ii)]  $\gamma\cap M\neq\phi.$ In this case,\\
            $\#\alpha=1\\
            \#\beta=(2^{|M|}-1)\times2^{m-|M|}+(2^{|N|}-1)\times 2^{m-|N|}-(2^{|M\cup N|}-1)\times2^{m-|M\cup N|}\\
            \#\gamma=2^{m-|M|}-1$.
        \end{itemize}
        Therefore, $\#(\alpha,\beta,\gamma)=2^{2m-|M|}-2^{2m-|M|-|N|}-2^{m}-2^{m-|M|}+2^{m-|N|}+2^{m-|M\cup N|}.$

        \item If $\varphi(\beta+\gamma|M)=0$ and $\varphi(\beta|N)=1,$ then $wt(c_D^{(2)}(\alpha,\beta, \gamma))=\frac{|D|}{2}.$\\
        \begin{itemize}
            \item [(i)] $\gamma\cap M=\phi.$
            In this case,\\
                $\#\alpha=1\\
                \#\beta=(2^{|M\setminus N|}-1)\times2^{m-|M\cup N|}\\
                \#\gamma=2^{m-|M|}-1$.
            \item[(ii)]  $\gamma\cap M\neq\phi.$
            \begin{itemize}
                \item[(a)] $\beta \cap M=\phi$.       
                In this case,\\
                    $\#\alpha=1\\
                    \#\beta=2^{m-|M\cup N|}-1\\
                    \#\gamma=(2^{|M|}-1)\times 2^{m-|M|}$.
                \item[(b)]$\beta \cap M\neq \phi$. In this case,\\
                    $\#\alpha=1\\
                    \#\beta=(2^{|M\setminus N|}-1)\times2^{m-|M\cup N|}\\
                    \#\gamma=(2^{|M|}-2)\times 2^{m-|M|}$.
            \end{itemize}
        \end{itemize}
        Therefore, $\#(\alpha,\beta,\gamma)=2^{2m-|N|}-2^{2m-|M|-|N|}-2^{m}+2^{m-|M|}-2^{m-|N|}+2^{m-|M\cup N|}.$

        \item If $\varphi(\beta+\gamma|M)=1$ and $\varphi(\beta|N)=1,$ then $wt(c_D^{(2)}(\alpha,\beta, \gamma))=\frac{|D|}{2}-(2^m-2^{|L|})\times 2^{|M|+|N|-1}.$
        \begin{itemize}
            \item [(i)] $\gamma\cap M=\phi.$
            In this case,\\
                $\#\alpha=1\\
                \#\beta=2^{m-|M\cup N|}-1\\
                \#\gamma=2^{m-|M|}-2$.
            \item[(ii)]  $\gamma\cap M\neq\phi.$  In this case,\\
                $\#\alpha=1\\
                \#\beta=(2^{|M\setminus N|}-1)\times2^{m-|M\cup N|}\\
                \#\gamma=2^{m-|M|}-1$.
        \end{itemize}
        Therefore, $\#(\alpha,\beta,\gamma)=2^{2m-|M|-|N|}-2^{m-|M\cup N|}-2^{m-|M|}-2^{m-|N|}+2.$
    \end{itemize}

\noindent
    Case (8): $\alpha\neq 0, \beta+\gamma\neq 0, \beta\neq0.$ In this case,
    \begin{equation*}
        wt(c_D^{(2)}(\alpha,\beta,\gamma))=\frac{|D|}{2}-2^{|L|+|M|+|N|-1}\varphi(\alpha|L)\varphi(\beta+\gamma|M)\varphi(\beta|N).
    \end{equation*}
    Subcase (1): $\gamma=0$
    \begin{itemize}
        \item If $\varphi(\alpha|L)=0,\,\varphi(\beta+\gamma|M)=0$ and $\varphi(\beta|N)=0,$ then $wt(c_D^{(2)}(\alpha,\beta, \gamma))=\frac{|D|}{2}.$
        In this case,\\
            $\#\alpha=(2^{|L|}-1)\times 2^{m-|L|}\\
            \#\beta=(2^{|M|}-1)\times2^{m-|M|}+(2^{|N|}-1)\times 2^{m-|N|}-(2^{|M\cup N|}-1)\times2^{m-|M\cup N|}\\
            \#\gamma=1$.\\
        Therefore, $\#(\alpha,\beta,\gamma)=2^{2m}-2^{2m-|M|}-2^{2m-|N|}+2^{2m-|M\cup N|}-2^{2m-|L|}+2^{2m-|M|-|L|}+2^{2m-|L|-|N|}-2^{2m-|L|-|M\cup N|}.$

         \item If $\varphi(\alpha|L)=1,\,\varphi(\beta+\gamma|M)=0$ and $\varphi(\beta|N)=0,$ then $wt(c_D^{(2)}(\alpha,\beta, \gamma))=\frac{|D|}{2}.$
        In this case,\\
            $\#\alpha=2^{m-|L|}-1\\
            \#\beta=(2^{|M|}-1)\times2^{m-|M|}+(2^{|N|}-1)\times 2^{m-|N|}-(2^{|M\cup N|}-1)\times2^{m-|M\cup N|}\\
            \#\gamma=1$.\\
        Therefore, $\#(\alpha,\beta,\gamma)=2^{2m-|L|}-2^{2m-|L|-|M|}-2^{2m-|L|-|N|}+2^{2m-|L|-|M\cup N|}-2^m+2^{m-|M|}+2^{m-|N|}-2^{m-|M\cup N|}.$

        \item If $\varphi(\alpha|L)=0,\,\varphi(\beta+\gamma|M)=1$ and $\varphi(\beta|N)=0,$ then $wt(c_D^{(2)}(\alpha,\beta, \gamma))=\frac{|D|}{2}.$
        In this case,\\
            $\#\alpha=(2^{|L|}-1)\times 2^{m-|L|}\\
            \#\beta=(2^{|N\setminus M|}-1)\times2^{m-|M\cup N|}\\
            \#\gamma=1$.\\
        Therefore, $\#(\alpha,\beta,\gamma)=2^{2m-|M|}-2^{2m-|M\cup N|}-2^{2m-|L|-|M|}+2^{2m-|L|-|M\cup N|}.$

         \item If $\varphi(\alpha|L)=0,\,\varphi(\beta+\gamma|M)=0$ and $\varphi(\beta|N)=1,$ then $wt(c_D^{(2)}(\alpha,\beta, \gamma))=\frac{|D|}{2}.$
        In this case,\\
            $\#\alpha=(2^{|L|}-1)\times 2^{m-|L|}\\
            \#\beta=(2^{|M\setminus N|}-1)\times2^{m-|M\cup N|}\\
            \#\gamma=1$.\\
        Therefore, $\#(\alpha,\beta,\gamma)=2^{2m-|N|}-2^{2m-|M\cup N|}-2^{2m-|L|-|N|}+2^{2m-|L|-|M\cup N|}.$

     \item If $\varphi(\alpha|L)=1, \,\varphi(\beta+\gamma|M)=1$ and $\varphi(\beta|N)=0,$ then $wt(c_D^{(2)}(\alpha,\beta, \gamma))=\frac{|D|}{2}.$
        In this case,\\
            $\#\alpha=2^{m-|L|}-1\\
            \#\beta=(2^{|N\setminus M|}-1)\times2^{m-|M\cup N|}\\
            \#\gamma=1$.\\
        Therefore, $\#(\alpha,\beta,\gamma)=2^{2m-|L|-|M|}-2^{2m-|L|-|M\cup N|}-2^{m-|M|}+2^{m-|M\cup N|}.$

    \item If $\varphi(\alpha|L)=1, \,\varphi(\beta+\gamma|M)=0$ and $\varphi(\beta|N)=1,$ then $wt(c_D^{(2)}(\alpha,\beta, \gamma))=\frac{|D|}{2}.$
        In this case,\\
            $\#\alpha=2^{m-|L|}-1\\
            \#\beta=(2^{|M\setminus N|}-1)\times2^{m-|M\cup N|}\\
            \#\gamma=1$.\\
        Therefore, $\#(\alpha,\beta,\gamma)=2^{2m-|L|-|N|}-2^{2m-|L|-|M\cup N|}-2^{m-|N|}+2^{m-|M\cup N|}.$

        \item If $\varphi(\alpha|L)=0, \,\varphi(\beta+\gamma|M)=1$ and $\varphi(\beta|N)=1,$ then $wt(c_D^{(2)}(\alpha,\beta, \gamma))=\frac{|D|}{2}.$
        In this case,\\
            $\#\alpha=(2^{|L|}-1)\times 2^{m-|L|}\\
            \#\beta=2^{m-|M\cup N|}-1\\
            \#\gamma=1$.\\
        Therefore, $\#(\alpha,\beta,\gamma)=2^{2m-|M\cup N|}-2^{2m-|L|-|M\cup N|}-2^{m}+2^{m-|L|}.$

        \item If $\varphi(\alpha|L)=1, \,\varphi(\beta+\gamma|M)=1$ and $\varphi(\beta|N)=1,$ then $wt(c_D^{(2)}(\alpha,\beta,\gamma))=\frac{|D|}{2}-2^{|L|+|M|+|N|-1}.$
        In this case,\\
            $\#\alpha=2^{m-|L|}-1\\
            \#\beta=2^{m-|M\cup N|}-1\\
            \#\gamma=1$.\\
        Therefore, $\#(\alpha,\beta,\gamma)=2^{2m-|L|-|M\cup N|}-2^{m-|L|}-2^{m-|M\cup N|}+1.$
        
    \end{itemize}
Subcase (2): $\gamma\neq 0,\,\gamma\neq \beta$
    \begin{itemize}
        \item If $\varphi(\alpha|L)=0,\,\varphi(\beta+\gamma|M)=0$ and $\varphi(\beta|N)=0,$ then $wt(c_D^{(2)}(\alpha,\beta, \gamma))=\frac{|D|}{2}.$
        \begin{itemize}
            \item [(i)] $\gamma\cap M=\phi.$
            In this case,\\
                $\#\alpha=(2^{|L|}-1)\times 2^{m-|L|}\\
                \#\beta=(2^{|M|}-1)\times2^{m-|M|}+(2^{|N|}-1)\times 2^{m-|N|}-(2^{|M\cup N|}-1)\times2^{m-|M\cup N|}\\
                \#\gamma=2^{m-|M|}-1$.
            \item[(ii)]  $\gamma\cap M\neq\phi.$
            \begin{itemize}
                \item[(a)] $\beta \cap M=\phi$.      
                In this case,\\
                    $\#\alpha=(2^{|L|}-1)\times 2^{m-|L|}\\
                    \#\beta=(2^{|N\setminus M|}-1)\times2^{m-|M\cup N|}\\
                    \#\gamma=(2^{|M|}-1)\times 2^{m-|M|}$.
                \item[(b)]$\beta \cap M\neq \phi$.
                In this case,\\
                    $\#\alpha=(2^{|L|}-1)\times 2^{m-|L|}\\
                    \#\beta=(2^{|M|}-1)\times2^{m-|M|}+(2^{|N|}-1)\times 2^{m-|N|}-(2^{|M\cup N|}-1)\times2^{m-|M\cup N|}\\
                    \#\gamma=(2^{|M|}-2)\times 2^{m-|M|}$.
            \end{itemize}
        \end{itemize}
        Therefore, $\#(\alpha,\beta,\gamma)=2^{3m}-2^{3m-|L|}-2^{3m-|M|}-2^{3m-|N|}+2^{3m-|L|-|M|}+2^{3m-|M|-|N|}+2^{3m-|L|-|N|}-2^{3m-|L|-|M|-|N|}-2^{2m}+2^{2m-|L|}+2^{2m-|M|}+2^{2m-|N|}-2^{2m-|M\cup N|}-2^{2m-|L|-|M|}-2^{2m-|L|-|N|}+2^{2m-|L|-|M\cup N|}.$

        \item If $\varphi(\alpha|L)=1,\,\varphi(\beta+\gamma|M)=0$ and $\varphi(\beta|N)=0,$ then $wt(c_D^{(2)}(\alpha,\beta, \gamma))=\frac{|D|}{2}.$
        \begin{itemize}
            \item [(i)] $\gamma\cap M=\phi.$
            In this case,\\
                $\#\alpha=2^{m-|L|}-1\\
                \#\beta=(2^{|M|}-1)\times2^{m-|M|}+(2^{|N|}-1)\times 2^{m-|N|}-(2^{|M\cup N|}-1)\times2^{m-|M\cup N|}\\
                \#\gamma=2^{m-|M|}-1$.
            \item[(ii)]  $\gamma\cap M\neq\phi.$
            \begin{itemize}
                \item[(a)] $\beta \cap M=\phi$.     
                In this case,\\
                $\#\alpha=2^{m-|L|}-1\\
                    \#\beta=(2^{|N\setminus M|}-1)\times2^{m-|M\cup N|}\\
                    \#\gamma=(2^{|M|}-1)\times 2^{m-|M|}$.
                \item[(b)]$\beta \cap M\neq \phi$. In this case,\\
                $\#\alpha=2^{m-|L|}-1\\
                    \#\beta=(2^{|M|}-1)\times2^{m-|M|}+(2^{|N|}-1)\times 2^{m-|N|}-(2^{|M\cup N|}-1)\times2^{m-|M\cup N|}\\
                    \#\gamma=(2^{|M|}-2)\times 2^{m-|M|}$.
            \end{itemize}
        \end{itemize}
        Therefore, $\#(\alpha,\beta,\gamma)=2^{3m-|L|}-2^{3m-|L|-|M|}-2^{3m-|L|-|N|}+2^{3m-|L|-|M|-|N|}-2^{2m}-2^{2m-|L|}+2^{2m-|M|}+2^{2m-|N|}+2^{2m-|L|-|M|}+2^{2m-|L|-|N|}-2^{2m-|M|-|N|}-2^{2m-|L|-|M\cup N|}+2^m-2^{m-|M|}-2^{m-|N|}+2^{m-|M\cup N|}.$

        \item If $\varphi(\alpha|L)=0,\,\varphi(\beta+\gamma|M)=1$ and $\varphi(\beta|N)=0,$ then $wt(c_D^{(2)}(\alpha,\beta, \gamma))=\frac{|D|}{2}.$
        \begin{itemize}
            \item [(i)] $\gamma\cap M=\phi.$
            In this case,\\
            $\#\alpha=(2^{|L|}-1)\times 2^{m-|L|}\\
                \#\beta=(2^{|N\setminus M|}-1)\times 2^{m-|M\cup N|}\\
                \#\gamma=2^{m-|M|}-2$.
            \item[(ii)]  $\gamma\cap M\neq\phi.$       
                In this case,\\
                $\#\alpha=(2^{|L|}-1)\times 2^{m-|L|}\\
                    \#\beta=(2^{|M|}-1)\times2^{m-|M|}+(2^{|N|}-1)\times 2^{m-|N|}-(2^{|M\cup N|}-1)\times2^{m-|M\cup N|}\\
                    \#\gamma=2^{m-|M|}-1$.
        \end{itemize}
        Therefore, $\#(\alpha,\beta,\gamma)=2^{3m-|M|}-2^{3m-|L|-|M|}-2^{3m-|M|-|N|}+2^{3m-|L|-|M|-|N|}-2^{2m}+2^{2m-|L|}-2^{2m-|M|}+2^{2m-|N|}+2^{2m-|M\cup N|}+2^{2m-|L|-|M|}-2^{2m-|L|-|N|}-2^{2m-|L|-|M\cup N|}.$

        \item If $\varphi(\alpha|L)=0,\,\varphi(\beta+\gamma|M)=0$ and $\varphi(\beta|N)=1,$ then $wt(c_D^{(2)}(\alpha,\beta, \gamma))=\frac{|D|}{2}.$
        \begin{itemize}
            \item [(i)] $\gamma\cap M=\phi.$
            In this case,\\
            $\#\alpha=(2^{|L|}-1)\times 2^{m-|L|}\\
                \#\beta=(2^{|M\setminus N|}-1)\times2^{m-|M\cup N|}\\
                \#\gamma=2^{m-|M|}-1$.
            \item[(ii)]  $\gamma\cap M\neq\phi.$
            \begin{itemize}
                \item[(a)] $\beta \cap M=\phi$.     
                In this case,\\
                $\#\alpha=(2^{|L|}-1)\times 2^{m-|L|}\\
                    \#\beta=2^{m-|M\cup N|}-1\\
                    \#\gamma=(2^{|M|}-1)\times 2^{m-|M|}$.
                \item[(b)]$\beta \cap M\neq \phi$. In this case,\\
                $\#\alpha=(2^{|L|}-1)\times 2^{m-|L|}\\
                    \#\beta=(2^{|M\setminus N|}-1)\times2^{m-|M\cup N|}\\
                    \#\gamma=(2^{|M|}-2)\times 2^{m-|M|}$.
            \end{itemize}
        \end{itemize}
        Therefore, $\#(\alpha,\beta,\gamma)=2^{3m-|N|}-2^{3m-|L|-|N|}-2^{3m-|M|-|N|}+2^{3m-|L|-|M|-|N|}-2^{2m}+2^{2m-|L|}+2^{2m-|M|}-2^{2m-|N|}+2^{2m-|M\cup N|}-2^{2m-|L|-|M|}+2^{2m-|L|-|N|}-2^{2m-|L|-|M\cup N|}.$

        \item If $\varphi(\alpha|L)=1,\,\varphi(\beta+\gamma|M)=1$ and $\varphi(\beta|N)=0,$ then $wt(c_D^{(2)}(\alpha,\beta, \gamma))=\frac{|D|}{2}.$
        \begin{itemize}
            \item [(i)] $\gamma\cap M=\phi.$
            In this case,\\
            $\#\alpha=2^{m-|L|}-1\\
                \#\beta=(2^{|N\setminus M|}-1)\times 2^{m-|M\cup N|}\\
                \#\gamma=2^{m-|M|}-2$.
            \item[(ii)]  $\gamma\cap M\neq\phi.$       
                In this case,\\
                $\#\alpha=2^{m-|L|}-1\\
                    \#\beta=(2^{|M|}-1)\times2^{m-|M|}+(2^{|N|}-1)\times 2^{m-|N|}-(2^{|M\cup N|}-1)\times2^{m-|M\cup N|}\\
                    \#\gamma=2^{m-|M|}-1$.
        \end{itemize}
        Therefore, $\#(\alpha,\beta,\gamma)=2^{3m-|L|-|M|}-2^{3m-|L|-|M|-|N|}-2^{2m-|L|}-2^{2m-|M|}-2^{2m-|L|-|M|}+2^{2m-|L|-|N|}+2^{2m-|M|-|N|}+2^{2m-|L|-|M\cup N|}+2^m+2^{m-|M|}-2^{m-|N|}-2^{m-|M\cup N|}.$

        \item If $\varphi(\alpha|L)=1,\,\varphi(\beta+\gamma|M)=0$ and $\varphi(\beta|N)=1,$ then $wt(c_D^{(2)}(\alpha,\beta, \gamma))=\frac{|D|}{2}.$
        \begin{itemize}
            \item [(i)] $\gamma\cap M=\phi.$
            In this case,\\
            $\#\alpha=2^{m-|L|}-1\\
                \#\beta=(2^{|M\setminus N|}-1)\times2^{m-|M\cup N|}\\
                \#\gamma=2^{m-|M|}-1$.
            \item[(ii)]  $\gamma\cap M\neq\phi.$
            \begin{itemize}
                \item[(a)] $\beta \cap M=\phi$.       
                In this case,\\
                $\#\alpha=2^{m-|L|}-1\\
                    \#\beta=2^{m-|M\cup N|}-1\\
                    \#\gamma=(2^{|M|}-1)\times 2^{m-|M|}$.
                \item[(b)]$\beta \cap M\neq \phi$. In this case,\\
                $\#\alpha=2^{m-|L|}-1\\
                    \#\beta=(2^{|M\setminus N|}-1)\times2^{m-|M\cup N|}\\
                    \#\gamma=(2^{|M|}-2)\times 2^{m-|M|}$.
            \end{itemize}
        \end{itemize}
        Therefore, $\#(\alpha,\beta,\gamma)=2^{3m-|L|-|N|}-2^{3m-|L|-|M|-|N|}-2^{2m-|L|}-2^{2m-|N|}+2^{2m-|L|-|M|}-2^{2m-|L|-|N|}+2^{2m-|M|-|N|}+2^{2m-|L|-|M\cup N|}+2^m-2^{m-|M|}+2^{m-|N|}-2^{m-|M\cup N|}.$

        \item If $\varphi(\alpha|L)=0,\,\varphi(\beta+\gamma|M)=1$ and $\varphi(\beta|N)=1,$ then $wt(c_D^{(2)}(\alpha,\beta, \gamma))=\frac{|D|}{2}.$
        \begin{itemize}
            \item [(i)] $\gamma\cap M=\phi.$
            In this case,\\
                $\#\alpha=(2^{|L|}-1)\times 2^{m-|L|}\\
                \#\beta=2^{m-|M\cup N|}-1\\
                \#\gamma=2^{m-|M|}-2$.
            \item[(ii)]  $\gamma\cap M\neq\phi.$       
                In this case,\\
                $\#\alpha=(2^{|L|}-1)\times 2^{m-|L|}\\
                    \#\beta=(2^{|M\setminus N|}-1)\times2^{m-|M\cup N|}\\
                    \#\gamma=2^{m-|M|}-1$.
        \end{itemize}
        Therefore, $\#(\alpha,\beta,\gamma)=2^{3m-|M|-|N|}-2^{3m-|L|-|M|-|N|}-2^{2m-|M|}-2^{2m-|N|}-2^{2m-|M\cup N|}+2^{2m-|L|-|M|}+2^{2m-|L|-|N|}+2^{2m-|L|-|M\cup N|}+2^{m+1}-2^{m-|L|+1}.$

        \item If $\varphi(\alpha|L)=1,\,\varphi(\beta+\gamma|M)=1$ and $\varphi(\beta|N)=1,$ then $wt(c_D^{(2)}(\alpha,\beta,\gamma))=\frac{|D|}{2}-2^{|L|+|M|+|N|-1}.$
        \begin{itemize}
            \item [(i)] $\gamma\cap M=\phi.$
            In this case,\\
            $\#\alpha=2^{m-|L|}-1\\
                \#\beta=2^{m-|M\cup N|}-1\\
                \#\gamma=2^{m-|M|}-2$.
            \item[(ii)]  $\gamma\cap M\neq\phi.$       
                In this case,\\
                    $\#\alpha=2^{m-|L|}-1,
                    \#\beta=(2^{|M\setminus N|}-1)\times2^{m-|M\cup N|}, 
                    \#\gamma=2^{m-|M|}-1$.
        \end{itemize}
        Therefore, $\#(\alpha,\beta,\gamma)=2^{3m-|L|-|M|-|N|}-2^{2m-|L|-|M|}-2^{2m-|L|-|N|}-2^{2m-|M|-|N|}-2^{2m-|L|-|M\cup N|}+2^{m-|M|}+2^{m-|N|}+2^{m-|M\cup N|}+2^{m-|L|+1}-2.$
    \end{itemize}
    Based on above computations, we obtain Table \ref{tab:8}.\\
     Observe that $|\ker(c_D^{(2)})|=1$ by case (1) and so by the first isomorphism theorem, we have
     \begin{equation*}
    |\mathcal{C}_D^{(2)}|=\frac{|\mathcal{R}_2^m|}{|\ker(c_D^{(2)})|}=2^{3m}.
     \end{equation*}
Hence, $\dim \mathcal{C}_D^{(2)}=3m.$ One obtains minimum distance based on Table \ref{tab:8}. This proves (8).
\par
We now prove the distance-optimal condition for part (5). Set
$$
n:=(2^m-2^{|L|})\times (2^m-2^{|M|})\times 2^{|N|}, k:=2m+|N|, d:=(2^m-2^{|L|}-2^{|M|})\times 2^{m+|N|-1}.
$$
\begin{align*}
    & \sum_{i=0}^{k-1} \left\lceil \frac{d+1}{2^i}\right\rceil\\
    &=\sum_{i=0}^{2m+|N|-1}\left\lceil \frac{2^{2m+|N|-1}-2^{m+|M|+|N|-1}-2^{m+|L|+|N|-1}+1}{2^i}\right\rceil\\
    &\ge 2^{2m+|N|}-1+\sum_{i=0}^{2m+|N|-1}\left\lceil \frac{1}{2^i}\right\rceil-\sum_{i=0}^{2m+|N|-1}\left\lfloor\frac{2^{m+|M|+|N|-1}+2^{m+|L|+|N|-1}}{2^i} \right\rfloor-1\\
    &\geq 2^{2m+|N|}-1+2m+|N|-\sum_{i=0}^{2m+|N|-1}\left\lfloor\frac{2^{m+|M|+|N|-1}}{2^i} \right\rfloor-\sum_{i=0}^{2m+|N|-1}\left\lfloor\frac{2^{m+|L|+|N|-1}}{2^i} \right\rfloor-2\\
    &= 2^{2m+|N|}-1+2m+|N|-2^{m+|M|+|N|}-2^{m+|L|+|N|}\\
    & \ge n+1 \iff 2(m-1)+|N|\ge 2^{|L|+|M|+|N|}. 
\end{align*}
\qed
\end{proof}
\begin{table}[H]
    \centering
\begin{tabular}{c|c} 
  \hline
  Hamming Weight & Number of codewords\\ 
  \hline
  $0$ & $1$\\
  \hline
  $2^{|L|+|M|+|N|-1}$ & $2^{|L|+|M|+|N|}-1$\\
  \hline
\end{tabular}
\caption{}
\label{tab:1}
\end{table}

\begin{table}[H]
    \centering
\begin{tabular}{c|c} 
  \hline
  Hamming Weight & Number of codewords\\ 
  \hline
  $0$ & $1$\\
  \hline
  $(2^m-2^{|L|})\times 2^{|M|+|N|-1}$ & $2^{m+|M|+|N|}-2^{m-|L|}$\\
  \hline
  $2^{m+|M|+|N|-1}$ & $2^{m-|L|}-1$\\
  \hline
\end{tabular}
\caption{}
\label{tab:2}
\end{table}

\begin{table}[H]
    \centering
\begin{tabular}{c|c} 
  \hline
  Hamming Weight & Number of codewords\\ 
  \hline
  $0$ & $1$\\
  \hline
  $(2^m-2^{|M|})\times 2^{|L|+|N|-1}$ & $2^{m+|L|+|N|}-2^{m-|M|}$\\
  \hline
  $2^{m+|L|+|N|-1}$ & $2^{m-|M|}-1$\\
  \hline
\end{tabular}
\caption{}
\label{tab:3}
\end{table}

\begin{table}[H]
    \centering
\begin{tabular}{ c|c } 
  \hline
  Hamming Weight & Number of codewords\\ 
  \hline
  $0$ & $1$\\
  \hline
  $(2^m-2^{|N|})\times 2^{|L|+|M|-1}$ & $2^{m+|L|+|M|}-2^{m-|N|}$\\
  \hline
  $2^{m+|L|+|M|-1}$ & $2^{m-|N|}-1$\\
  \hline
\end{tabular}
\caption{}
\label{tab:4}
\end{table}

\begin{table}[H]
    \centering
\begin{tabular}{ c|c } 
  \hline
  Hamming Weight & Number of codewords\\ 
  \hline
  $0$ & $1$\\
  \hline
  $(2^m-2^{|L|})\times (2^m-2^{|M|})\times 2^{|N|-1}$ & $2^{2m+|N|}-2^{2m-|L|-|M|}$\\
  \hline
  $(2^m-2^{|M|})\times 2^{m+|N|-1}$ & $2^{m-|L|}-1$\\
  \hline
  $(2^m-2^{|L|})\times 2^{m+|N|-1}$ & $2^{m-|M|}-1$\\
  \hline
  $(2^m-2^{|L|}-2^{|M|})\times 2^{m+|N|-1}$ & $2^{2m-|L|-|M|}-2^{m-|L|}-2^{m-|M|}+1$\\
  \hline
\end{tabular}
\caption{}
\label{tab:5}
\end{table}

\begin{table}[H]
    \centering
\begin{tabular}{ c|c } 
  \hline
  Hamming Weight & Number of codewords\\ 
  \hline
  $0$ & $1$\\
  \hline
  $(2^m-2^{|L|})\times (2^m-2^{|N|})\times 2^{|M|-1}$ & $2^{2m+|M|}-2^{2m-|L|-|N|}$\\
  \hline
  $(2^m-2^{|N|})\times 2^{m+|M|-1}$ & $2^{m-|L|}-1$\\
  \hline
  $(2^m-2^{|L|})\times 2^{m+|M|-1}$ & $2^{m-|N|}-1$\\
  \hline
  $(2^m-2^{|L|}-2^{|N|})\times 2^{m+|M|-1}$ & $2^{2m-|L|-|N|}-2^{m-|L|}-2^{m-|N|}+1$\\
  \hline
\end{tabular}
\caption{}
\label{tab:6}
\end{table}

\begin{table}[H]
    \centering
\begin{tabular}{ c|c } 
  \hline
  Hamming Weight & Number of codewords\\ 
  \hline
  $0$ & $1$\\
  \hline
  $(2^m-2^{|M|})\times (2^m-2^{|N|})\times 2^{|L|-1}$ & $2^{2m+|L|}-2^{2m-|M|-|N|}$\\
  \hline
  $(2^m-2^{|M|})\times 2^{m+|L|-1}$ & $2^{m-|N|}-1$\\
  \hline
  $(2^m-2^{|N|})\times 2^{m+|L|-1}$ & $2^{m-|M|}-1$\\
  \hline
  $(2^m-2^{|M|}-2^{|N|})\times 2^{m+|L|-1}$ & $2^{2m-|M|-|N|}-2^{m-|N|}-2^{m-|M|}+1$\\
  \hline
\end{tabular}
\caption{}
\label{tab:7}
\end{table}

\begin{table}[H]
    \centering
\scalebox{0.8}{\begin{tabular}{ c|c } 
  \hline
  Hamming Weight & Number of codewords\\ 
  \hline
  $0$ & $1$\\
  \hline
  $2^{-1}\times(2^m-2^{|L|})\times (2^m-2^{|M|})\times (2^m-2^{|N|})$ & $2^{3m}-2^{3m-|L|-|M|-|N|}$\\
  \hline
  $2^{m-1}\times(2^m-2^{|M|})\times (2^m-2^{|N|})$ & $2^{m-|L|}-1$\\
  \hline
  $2^{m-1}\times(2^m-2^{|L|})\times (2^m-2^{|N|})$ & $2^{m-|M|}-1$\\
  \hline
  $2^{m-1}\times(2^m-2^{|L|})\times (2^m-2^{|M|})$ & $2^{m-|N|}-1$\\
  \hline
  $2^{m-1}\times(2^m-2^{|L|})\times (2^m-2^{|M|}-2^{|N|})$ & $2^{2m-|M|-|N|}-2^{m-|M|}-2^{m-|N|}+1$\\
  \hline
  $2^{m-1}\times(2^m-2^{|M|})\times (2^m-2^{|L|}-2^{|N|})$ & $2^{2m-|L|-|N|}-2^{m-|L|}-2^{m-|N|}+1$\\
  \hline
  $2^{m-1}\times(2^m-2^{|N|})\times (2^m-2^{|L|}-2^{|M|})$ & $2^{2m-|L|-|M|}-2^{m-|L|}-2^{m-|M|}+1$\\
  \hline
  $2^{-1}\times \left((2^m-2^{|L|})\times(2^m-2^{|M|})\times(2^m-2^{|N|})+2^{|L|+|M|+|N|}\right)$ & $2^{3m-|L|-|M|-|N|}-2^{2m-|M|-|N|}-2^{2m-|L|-|N|}$\\
    & $-2^{2m-|L|-|M|}+2^{m-|L|}+2^{m-|M|}+2^{m-|N|}-1$\\
  \hline
\end{tabular}}
\caption{}
\label{tab:8}
\end{table}

\begin{table}[H]
    \centering
\begin{tabular}{ c|c } 
  \hline
  Hamming Weight & Number of codewords\\ 
  \hline
  $0$ & $1$\\
  \hline
  $2^{3m-1}$ & $2^{3m-|L|-|M|-|N|}-1$\\
  \hline
  $2^{3m-1}-2^{|L|+|M|+|N|-1}$ & $2^{3m}-2^{3m-|L|-|M|-|N|}$\\
  \hline
\end{tabular}
\caption{}
\label{tab:9}
\end{table}
\begin{theorem}\label{minimality condition}
    Let $m\in \mathbb{N}$ and let $L, M, N\subseteq [m].$ Assume $\mathcal{C}_{D}^{(2)}$ as in Theorem \ref{main theorem}. Then the sufficient condition for $\mathcal{C}_D^{(2)}$ to be minimal and self-orthogonal is given in Table \ref{tab:10} for each $D.$  
\end{theorem}
\begin{table}[H]
    \centering
\begin{tabular}{c|c|c|c } 
  \hline
  Sl.No. & $D$ as in & Minimality condition & Self-orthogonality condition\\ 
  \hline
  1& Theorem \ref{main theorem}(\ref{main theorem 1}) & Always minimal & $|L|+|M|+|N|\ge 3$\\
  \hline
  2& Theorem \ref{main theorem}(\ref{main theorem 2})& $|L|\leq m-2$ & $|M|+|N|\ge 3$\\
  \hline
  3& Theorem \ref{main theorem}(\ref{main theorem 3})& $|M|\leq m-2$ & $|L|+|N|\ge 3$\\
  \hline
  4& Theorem \ref{main theorem}(\ref{main theorem 4})& $|N|\leq m-2$ & $|L|+|M|\ge 3$\\
  \hline
  5& Theorem \ref{main theorem}(\ref{main theorem 5})& $\max \{|L|, |M|\}\leq m-2$ & $|N|\ge 3$\\
  \hline
  6& Theorem \ref{main theorem}(\ref{main theorem 6})& $\max \{|L|, |N|\}\leq m-2$ & $|M|\ge 3$\\
  \hline
  7& Theorem \ref{main theorem}(\ref{main theorem 7})& $\max \{|M|, |N|\}\leq m-2$ & $|L|\ge 3$\\
  \hline
  8& Theorem \ref{main theorem}(\ref{main theorem 8})& $\max \{|L|, |M|, |N|\}\leq m-2$ & $L,M,N \neq \phi$\\
  \hline
  9& Theorem \ref{main theorem}(\ref{main theorem 9})& $|L|+|M|+|N|\leq 3m-2$ & $|L|+|M|+|N|\ge 3$ \\
  \hline
\end{tabular}
\caption{Minimality and self-orthogonality conditions in Theorem \ref{minimality condition}}
\label{tab:10}
\end{table}
\begin{proof}
    Proving the self-orthogonality conditions are straight forward. We just derive the sufficient condition for minimality of Sl. No. 8 and the proofs of the other parts follows similarly.
    \par
     From Table \ref{tab:8}, if $s_1\le s_2\le s_3$ are such that  $\{s_1, s_2, s_3\}=\{|L|,|M|,|N|\},$ then
     \begin{align*}
         wt_{min}&=2^{m-1}\times (2^{m}-2^{s_1})\times (2^m-2^{s_2}-2^{s_3})\\
         wt_{max}&=2^{m-1}\times (2^{m}-2^{s_1})\times (2^m-2^{s_2}).
     \end{align*}
    Now, $\frac{wt_{min}}{wt_{max}}>\frac{1}{2} \iff 2^m>2^{s_2}+2^{s_3+1}.$ Since $s_2\le s_3,\, 2^m>2^{s_2}+2^{s_3+1}\impliedby 2^m>2^{s_3}+2^{s_3+1}\impliedby s_3\le m-2.$ Hence, $\mathcal{C}_D^{(2)}$ is minimal if $\max\{|L|,|M|,|N|\}\le m-2.$ \qed
\end{proof}
The following examples illustrate the above results.
\begin{example}
    If $m\ge 2, |L|\le m-2$ and  $|M|+|N|\ge 3,$ then the code obtained in Theorem \ref{main theorem}(\ref{main theorem 2}) is a minimal, self-orthogonal, distance-optimal $2$-weight binary linear code.
\end{example}
\begin{example}
    If $m=2^t, t\ge 2, |N|\ge 3$ and $|L|+|M|\le t-1,$ then the code obtained in Theorem \ref{main theorem}(\ref{main theorem 5}) is a minimal, self-orthogonal, distance-optimal $4$-weight binary linear code.
\end{example}

\section{Code comparison}\label{Section 5}
This section contains three tables that demonstrate the significance of our codes. In Table \ref{table:11}, we present some recent works on binary linear codes constructed with the help of simplicial complexes. For the reader's convenience, we list certain few-weight codes of this article having good parameters in Table \ref{table:12}. Table \ref{table:ListNewCodes} presents several numerical examples of binary codes which are obtained using Theorem \ref{main theorem}. We have verified with MAGMA software\cite{bosma1997magma} that all the codes in this table are optimal. Remark ``New'' in Table \ref{table:ListNewCodes} indicates that the corresponding codes are new in the sense that they are inequivalent to the currently Best-known linear codes. We believe that more new optimal linear codes with code length $>$ 256 can be obtained by using Theorem \ref{main theorem}, but we cannot verify that by using the current database \cite{codetable}.

\begin{table}[H]
	\centering
	
	\scalebox{0.7}{\begin{tabular}{|c|c|c|c|c|c|c|}
			\hline
			& $m$ &	$L$ & $M$  & $N$ & $[n,k,d]$ & Remark   \\
			\hline
		 	\multirow{3}{*}{Theorem \ref{main theorem} (\ref{main theorem 2})} & $m = 3$ & $(1, 0, 0)$ & $(1, 1, 0)$  & $(1, 1, 1)$ & $[192, 8, 96]$& $-$\\
			\cline{2-7}
			      & {$m = 4$}  &  $(1, 0, 0, 0)$  & $( 0, 0, 0, 0)$   & $(0, 1, 1, 1)$ &$[112, 7, 56]$ & $-$\\
			\cline{2-7}
			     &    $m = 5$   &  $(0, 1, 0, 0, 0)$   & $(1, 0, 0, 0, 0)$    & $(0, 0, 1, 0, 1)$ &   $[240, 8, 120]$ & $-$\\
			\hline 
			\multirow{18}{*}{Theorem \ref{main theorem} (\ref{main theorem 5})} & \multirow{2}{*}{$m = 2$} & $(0, 0)$ & $(0, 0)$  & $(1, 1)$ & $[36, 6, 16]$ & New\\
			\cline{3-7}
			    &  &  $(0, 0)$  & $( 1, 0)$   & $(0, 0)$ &$[6, 4, 2]$ & New\\
			\cline{2-7}
			       & \multirow{7}{*}{$m = 3$} & $(0, 0, 0)$ & $(0, 0, 0)$  & $(1, 0, 1)$ & $[196, 8, 96]$ & New\\
			\cline{3-7}
			   &  &  $(0, 0, 0)$  & $( 0, 0, 1)$   & $(0, 0, 0)$ &$[42, 6, 20]$ & New\\
			   \cline{3-7}
			   &  &  $(0, 0, 0)$  & $( 0, 1, 0)$   & $(1, 0, 0)$ &$[84, 7, 40]$ & New\\
			   \cline{3-7}
			   &  &  $(0, 0, 0)$  & $( 1, 0, 1)$   & $(0, 0, 0)$ &$[28, 6, 12]$ & New\\
			    \cline{3-7}
			   &  &  $(0, 1, 0)$  & $( 0, 0, 0)$   & $(0, 0, 0)$ &$[42, 6, 20]$ & New\\
			    \cline{3-7}
			   &  &  $(1, 0, 0)$  & $(0, 0, 0)$   & $(1, 0, 0)$ &$[84, 7, 40]$ & New\\
			    \cline{3-7}
			   &  &  $(0, 0, 1)$  & $(0, 1, 0)$   & $(0, 0, 0)$ &$[36, 6, 16]$ & New\\
			   \cline{2-7}
			   & \multirow{6}{*}{$m=4$}  &  $(0, 0, 0, 0)$  & $(0, 0, 0, 0)$   & $(0, 0, 0, 0)$ &$[225, 8, 112]$ & New\\
			   \cline{3-7}
			   &   &  $(0, 0, 0, 0)$  & $(1, 0, 0, 0)$   & $(0, 0, 0, 0)$ &$[210, 8, 104]$ & $-$\\
			   \cline{3-7}
			    &   &  $(0, 0, 0, 0)$  & $(0, 1, 0, 1)$   & $(0, 0, 0, 0)$ &$[180, 8, 88]$ & New\\
			    \cline{3-7}
			    &   &  $(1, 0, 0, 0)$  & $(0, 0, 0, 0)$   & $(0, 0, 0, 0)$ &$[210, 8, 104]$ & $-$\\
			    \cline{3-7}
			    &   &  $(0, 0, 1, 0)$  & $(0, 0, 0, 1)$   & $(0, 0, 0, 0)$ &$[196, 8, 96]$ & New\\
			    \cline{3-7}
			    &   &  $(0, 1, 1, 0)$  & $(0, 0, 0, 0)$   & $(0, 0, 0, 0)$ &$[180, 8, 88]$ & New\\
			\hline
			\multirow{3}{*}{Theorem \ref{main theorem} (\ref{main theorem 9})} & $m=2$ & $(1, 1)$ & $(1, 1)$ & $(0, 0)$ & $[48, 6, 24]$ & $-$\\
			\cline{2-7}
			  &   \multirow{2}{*}{$m=3$} & $(1, 1, 0)$ & $(1, 1, 1)$ & $(1, 1, 1)$ & $[256, 9, 128]$ & $-$\\
			  \cline{3-7}
			  &    & $(1, 1, 1)$ & $(0, 1, 1)$ & $(1, 1, 1)$ & $[256, 9, 128]$ & $-$\\
			  \hline
	\end{tabular}}
	\caption{Binary distance-optimal linear codes obtained from Theorem \ref{main theorem}}
	\label{table:ListNewCodes} 		
\end{table}

\begin{landscape}
\begin{table}[]
    \centering
		  \scalebox{0.6}{\begin{tabular}{|c|c|c|c|c|c|c|}
			\hline
			Reference & Result  & $[n,k,d]$-code & \#Weight & Distance-optimal& Minimal & Bound \\
			\hline
                \multirow{3}{*}{\cite{sagarI}}
                & Theorem 4.3 & $[(2^m-2^{|M|})\times 2^{|N|+1}, m, (2^m-2^{|M|})\times 2^{|N|}]$ & $2$ & Yes, if $1 \leq 2^{|N|+1}-1<|M|+|N|+1 \leq m$ & Yes, if $|M|\le m-2$ & -\\
                & Theorem 4.3 & $[2\times(2^m-2^{|M|})\times(2^m-2^{|N|}), m, (2^m-2^{|M|})\times(2^m-2^{|N|})]$ & $2$ & - & Yes, if $|M|\le m-2$ & -\\
                & Theorem 4.3 & $[2\times(2^{2m}-2^{|M|+|N|}), m, 2^{2m}-2^{|M|+|N|}]$ & $2$ & - & Yes, if $|M|+|N|\le 2m-2$&-\\
                \hline
                
                \multirow{6}{*}{\cite{sagarE}}
                & Theorem 4.4 & $[(2^m-2^{|M|})\times 2^{|N|}, m, (2^m-2^{|M|})\times 2^{|N|-1}]$ & $2$ & Yes, if $1 \leq 2^{|N|}-1<|M|+|N|\le m$ & Yes, if $|M|+|N|\ge 3$ &-\\
                & Theorem 4.4 & $[(2^m-2^{|M|})\times(2^m-2^{|N|}), m, (2^m-2^{|M|})\times(2^{m-1}-2^{|N|-1})]$ & $2$ & - & Yes, if $|M|\le m-2$ &-\\
                & Theorem 4.4 & $[2^{2m}-2^{|M|+|N|}, m, 2^{2m-1}-2^{|M|+|N|-1}]$ & $2$ & - & Yes, if $|M|+|N|\le 2m-2$ &-\\
                & Theorem 5.3 & $[2^{|M|+1}\times (2^m-2^{|N|}), m, (2^m-2^{|N|})\times 2^{|M|}]$ & $3$ & Yes, if $1 \leq 2^{|M|+1}-1<|M|+|N|+1\le m$ & Yes, if $|N|\le m-2$ &-\\
                & Theorem 5.3 & $[2\times(2^m-2^{|M|})\times(2^m-2^{|N|}), m, (2^m-2^{|M|})\times(2^{m}-2^{|N|})]$ & $3$ & - & Yes &-\\
                & Theorem 5.3 & $[2\times (2^{2m}-2^{|M|+|N|}), m, 2^{2m}-2^{|M|+|N|}]$ & $3$ & - & Yes, if $|M|+|N|\le 2m-2$ &-\\
                \hline
			\cite{sagar2023octanary} & Theorem 4.6 & $[2^{3m}-2^{|L|+|M|+|N|}, 3m, 2^{3m-1}- 2^{|L|+|M|+|N|-1}]$ & $2$ & Yes & Yes, if $ |L|+|M|+|N|\le 3m-2$ & Griesmer \\
                \hline

                \multirow{4}{*}{\cite{wu2022quaternary}}
                & Theorem 5.2 & $[2^{2m}-2^{|A|+|B|}, 2m, 2^{2m-1}-2^{|A|+|B|-1}]$ & $2$ & Yes & -& Griesmer \\
                & Proposition 5.3 & $[(2^{|A|}+2^{|B|}-2^{|A\cap B|})^2-1, 2\times |A\cup B|]$ & $\le 10$ & - & - &-\\
                & Proposition 5.5 & $[(2^{|A|}+2^{|B|}-2^{|A\cap B|})^2-1, (2^{|A|}+2^{|B|}-2^{|A\cap B|})^2-1-2\times |A\cup B|, 3]$ &-  & - & - & Sphere packing \\
                & Proposition 5.6 & $[4^m-(2^{|A|}+2^{|B|}-2^{|A\cap B|})^2, 2m]$ & $\le 11$ & - & -&-\\
                \hline

                \multirow{2}{*}{\cite{wu2020binary}}
                & Theorem 5 & $[2^{|A|}-2^{|B|}, |A|, 2^{|A|-1}-2^{|B|-1}]$ & $2$ & Yes & - & Griesmer \\
                & Theorem 6 & $[(2^{|A|}-2^{|B|}, 2^{|A|}-2^{|B|}-|A|, 3 \,\text{or}\, 4]$ & - & - & -&-\\
                \hline
                
                \multirow{5}{*}{\cite{hyun2019optimal}}
                & Theorem 4.1 & $[2^m-r-1, m, 2^{m-1}-r]$ & $2$ & Yes & -& Griesmer \\
                & Theorem 4.4 & $[2^m-2r-2, m, 2^{m-1}-2r-1]$ & $4$ & Yes & -& Griesmer \\
                & Theorem 4.7 & $[2^m-3r-3, m, 2^{m-1}-3r-2]$ & $5$ & Yes & -& Griesmer \\
                & Theorem 4.11 & $[2^m-r-1, m, 2^{m-1}-1]$ & $4$ & Yes & -& Griesmer \\
                & Theorem 4.14 & $[2^m, m, 2^{m-1}+r-1]$ & $5$ & Yes & -& Griesmer \\
                \hline

                \multirow{7}{*}{\cite{hyun2020infinite}}
                & Lemma 7 & $[2^m-2^{|A|}, m, 2^{m-1}-2^{|A|-1}]$ & $2$ & Yes & -& Griesmer \\
                & Example 10 & $[2^{m-1}, m, 4]$ & - & - & -& Sphere packing \\
                & Corollary 21 & $[2^m-2\times\sum_{i=1}^{s}2^{|A_i|-1}+s-1, m, 2^{m-1}\sum_{i=1}^{s}2^{|A_i|-1}]$ &-  & Yes & - & Griesmer \\
                &Lemma 26 & $[2^{ \vert A_1 \vert }+2^{ \vert A_2 \vert }-2,  \vert A_1\cup A_2 \vert , 2^{ \vert A_1 \vert -1}]$ & $3$ &- &- & - \\
                & Lemma 26 & $[2^{ \vert A_1 \vert }+2^{ \vert A_2 \vert }-2^{ \vert A_1\cap A_2 \vert }-1,  \vert A_1\cup A_2 \vert , 2^{ \vert A_1 \vert -1}]$ & $4$ & - & - &-\\
			& Theorem 27 & $[2^m-2^{ \vert A_1 \vert }-2^{ \vert A_2 \vert }+1, m, 2^{m-1}-2^{ \vert A_1 \vert -1}2^{ \vert A_2 \vert -1}]$ & $3$ or $4$ & Yes &- & Griesmer \\
			& Theorem 27 & $[2^m-2^{ \vert A_1 \vert }-2^{ \vert A_2 \vert }+2^{ \vert A_1\cap A_2 \vert }-1, m, 2^{m-1}-2^{ \vert A_1 \vert -1}-2^{ \vert A_2 \vert -1}]$ & $4$ or $5$ & Yes &- & -\\
            \hline
                
                \multirow{3}{*}{\cite{sagar2022certain}}
                & Theorem 5.2 & $[(2^m-2^{|L|})\times 2^{|M|+|N|}, m, (2^m-2^{|L|})\times 2^{|M|+|N|-1}]$ & $2$ & Yes, if $0\leq 2^{|M|+|N|}-1\leq |L|+|M|+|N| \leq m$ & Yes, if $|L|\leq m-2$ &Griesmer\\
                & Theorem 5.2 & $[(2^m-2^{|L|})\times(2^m-2^{|N|})\times 2^{|M|}, m, (2^m-2^{|L|})\times(2^m-2^{|N|})\times 2^{|M|-1}]$ & $2$ & - & Yes, if $|L|\le m-2$ &-\\
                & Theorem 5.2 & $[(2^m-2^{|L|})\times(2^m-2^{|M|})\times (2^{m}-2^{|N|}), m, (2^m-2^{|M|})\times(2^m-2^{|L|})\times (2^{m-1}-2^{|N|-1})]$ & $2$ & - & Yes, if $|L|\le m-2$ &-\\
                \hline  
            \end{tabular}}
        \caption{Certain binary linear codes from simplicial complexes before this article}
	\label{table:11} 		
\end{table}
\end{landscape}

\begin{landscape}

\begin{table}[]
	\centering
		\scalebox{0.65}{\begin{tabular}{|c|c|c|c|c|c|c|}
			\hline
			Result &	Defining set & Binary $[n,k,d]$-code & \#Weight & Distance-optimal & Minimal & Bound \\
            \hline
            Theorem \ref{main theorem}
            (\ref{main theorem 1}) & $\bm{e}_1\Delta_L+\bm{e}_2\Delta_M+\bm{e}_3\Delta_N$ & $[2^{|L|+|M|+|N|},|L|+|M|+|N|, 2^{|L|+|M|+|N|-1}]$ & 1 & Yes & Yes & Griesmer\\
		  \hline
            Theorem \ref{main theorem}
            (\ref{main theorem 2}) & $\bm{e}_1\Delta_L^c+\bm{e}_2\Delta_M+\bm{e}_3\Delta_N$	& $[(2^m-2^{|L|})\times 2^{|M|+|N|},m+|M|+|N|, (2^m-2^{|L|})\times 2^{|M|+|N|-1}]$ & 2 & Yes & Yes, if $|L|\le m-2$ & Griesmer\\
		  \hline
            Theorem \ref{main theorem}
            (\ref{main theorem 3}) & $\bm{e}_1\Delta_L+\bm{e}_2\Delta_M^c+\bm{e}_3\Delta_N$	& $[(2^m-2^{|M|})\times 2^{|L|+|N|},m+|L|+|N|, (2^m-2^{|M|})\times 2^{|L|+|N|-1}]$ & 2 & Yes & Yes, if $|M|\le m-2$& Griesmer\\
		  \hline
            Theorem \ref{main theorem}
            (\ref{main theorem 4}) & $\bm{e}_1\Delta_L+\bm{e}_2\Delta_M+\bm{e}_3\Delta_N^c$	& $[(2^m-2^{|N|})\times 2^{|L|+|M|},m+|L|+|M|, (2^m-2^{|N|})\times 2^{|L|+|M|-1}]$ & 2 & Yes & Yes, if $|N|\le m-2$& Griesmer\\
		  \hline
            Theorem \ref{main theorem}
            (\ref{main theorem 5}) & $\bm{e}_1\Delta_L^c+\bm{e}_2\Delta_M^c+\bm{e}_3\Delta_N$	& $[(2^m-2^{|L|})\times (2^m-2^{|M|})\times 2^{|N|}, 2m+|N|, (2^m-2^{|L|}-2^{|M|})\times 2^{m+|N|-1}]$ & 4 & Yes, if $2^{|L|+|M|+|N|}\le 2(m-1)+|N|$ & Yes, if $\max \{|L|, |M|\}\leq m-2$ &-\\
		  \hline
            Theorem \ref{main theorem}
            (\ref{main theorem 6}) & $\bm{e}_1\Delta_L^c+\bm{e}_2\Delta_M+\bm{e}_3\Delta_N^c$	& $[(2^m-2^{|L|})\times (2^m-2^{|N|})\times 2^{|M|}, 2m+|M|, (2^m-2^{|L|}-2^{|N|})\times 2^{m+|M|-1}]$ & 4 & Yes, if $2^{|L|+|M|+|N|}\le 2(m-1)+|M|$ & Yes, if $\max \{|L|, |N|\}\leq m-2$ &-\\
		  \hline
            Theorem \ref{main theorem}
            (\ref{main theorem 7})& $\bm{e}_1\Delta_L+\bm{e}_2\Delta_M^c+\bm{e}_3\Delta_N^c$	& $[(2^m-2^{|M|})\times (2^m-2^{|N|})\times 2^{|L|}, 2m+|L|, (2^m-2^{|M|}-2^{|N|})\times 2^{m+|L|-1}]$ & 4 & Yes, if $2^{|L|+|M|+|N|}\le 2(m-1)+|L|$ & Yes, if $\max \{|M|, |N|\}\leq m-2$ &-\\
		  \hline
            Theorem \ref{main theorem}
            (\ref{main theorem 8}) & $\bm{e}_1\Delta_L^c+\bm{e}_2\Delta_M^c+\bm{e}_3\Delta_N^c$	& $[(2^m-2^{|L|})\times (2^m-2^{|M|})\times (2^{M}-2^{|N|}), 3m]$ & 8 & - & Yes, if $\max \{|L|, |M|, |N|\}\leq m-2$ &-\\
		  \hline
            Theorem \ref{main theorem}
            (\ref{main theorem 9}) & $\mathcal{R}_2^m\setminus \bm{e}_1\Delta_L+\bm{e}_2\Delta_M+\bm{e}_3\Delta_N$	& $[(2^{3m}-2^{|L|+|M|+|N|}), 3m, 2^{3m-1}-2^{|L|+|M|+|N|-1}]$ & 2 & Yes & Yes, if $|L|+|M|+|N|\leq 3m-2$ &Griesmer\\
		  \hline
\end{tabular}}
		
	\caption{Binary linear codes from simplicial complexes in this article}
	\label{table:12}	
\end{table}
\end{landscape}

\section{Conclusion and Future Work}\label{Section 6}
In this article, we used an $\mathbb{F}_2$-valued trace of $\mathcal{R}_2=\mathbb{F}_2[x]/\langle x^3-x\rangle$ to study the binary subfield code of $\mathcal{C}_D$ over $\mathcal{R}_2,$ where the defining set $D$ is derived from a simplicial complex with one maximal element. The parameters and the Hamming weight distributions of $\mathcal{C}_D^{(2)}$ were determined for every defining set $D$. We also proved that some of these codes meet the Griesmer bound, and if the defining set is appropriately chosen, then the corresponding codes are distance-optimal. Furthermore, we produced a few infinite families of minimal, self-orthogonal and distance-optimal binary linear codes that are either $2$-weight or $4$-weight. In Table \ref{table:ListNewCodes}, we have displayed a few numerical examples of codes corresponding to Theorem \ref{main theorem} in which quite a few of them are new, that is, they are not equivalent to the Best-known linear codes.
\par
In \cite{sagar2022certain}, certain subfield-like codes of $\mathcal{C}_D$-codes over $\mathcal{R}_2$ are studied. It is also worth mentioning that the subfield codes of $\mathcal{C}_D$-codes over $\mathcal{R}_2$ obtained in this article possess better parameters compared to the parameters of the subfield-like codes in \cite{sagar2022certain}.
\par
Here, we considered simplicial complexes having only one maximal element. Working with two maximal elements becomes tedious, and hence, in future, one may try to find a different technique to obtain the weight distribution of $\mathcal{C}_D^{(2)},$ where the defining set is derived from a simplicial complex having multiple maximal elements.

\section*{Declarations}
\subsection*{Conflict of Interest}
All authors declare that they have no conflict of interest.

\section*{Acknowledgements}
The work of the first author was supported by Council of Scientific and Industrial Research (CSIR) India, under the grant no. 09/0086(13310)/2022-EMR-I. The third author thanks Indian Statistical Institute Delhi, India, for financial support during the research visit.

\bibliographystyle{abbrv}
\bibliography{Simplicial_Complex}

\end{document}